\documentclass[article, aps, 12pt, floatfix, longbibliography, a4paper]{revtex4-2}

\usepackage{graphicx}
\usepackage{epstopdf}
\usepackage{hyperref}

\usepackage{color,soul}
\usepackage{amsmath,bm}
\usepackage[utf8]{inputenc}
\usepackage{physics}
\usepackage{amsmath,bm}
\usepackage[normalem]{ulem}
\usepackage[utf8]{inputenc}
\usepackage{xcolor}
\sethlcolor{yellow}

\begin{document}

\title[]{Lattice-commensurate skyrmion texture in a centrosymmetric breathing kagome magnet}

\author{Max Hirschberger$^{1,2}$}\email{hirschberger@ap.t.u-tokyo.ac.jp}
\author{Bertalan G. Szigeti$^{3}$}
\author{Mamoun Hemmida$^{3}$}
\author{Moritz M. Hirschmann$^{2}$}
\author{Sebastian Esser$^{1}$}
\author{Hiroyuki Ohsumi$^{4}$}
\author{Yoshikazu Tanaka$^{4}$}
\author{Leonie Spitz$^{2,\dagger}$}
\author{Shang Gao$^{2,\S}$}
\author{Kamil K. Kolincio$^{2,5}$}
\author{Hajime Sagayama$^{6}$}
\author{Hironori Nakao$^{6}$}
\author{Yuichi Yamasaki$^{7}$}
\author{L{\'a}szl{\'o} Forr{\'o}$^{8}$}
\author{Hans-Albrecht Krug von Nidda$^{3}$}
\author{Istvan Kezsmarki$^{3}$}
\author{Taka-hisa Arima$^{2,9}$}
\author{Yoshinori Tokura$^{1,2,10}$}

\affiliation{$^{1}$Department of Applied Physics and Quantum-Phase Electronics Center, The University of Tokyo, Bunkyo-ku, Tokyo 113-8656, Japan}
\affiliation{$^{2}$RIKEN Center for Emergent Matter Science (CEMS), Wako, Saitama 351-0198, Japan}
\affiliation{$^{3}$Experimental Physics V, Center for Electronic Correlations and Magnetism, University of Augsburg, 86135 Augsburg, Germany}
\affiliation{$^{4}$RIKEN SPring-8 Center, 1-1-1 Kouto, Sayo, Hyogo 679-5148, Japan}
\affiliation{$^{5}$Faculty of Applied Physics and Mathematics, Gda{\'n}sk University of Technology, Narutowicza 11/12, 80-233 Gda{\'n}sk, Poland}
\affiliation{$^{6}$Institute of Materials Structure Science, High Energy Accelerator Research Organization, Tsukuba, Ibaraki 305-0801, Japan}
\affiliation{$^{7}$Research and Services Division of Materials Data and Integrated System (MaDIS), National Institute for Materials Science (NIMS), Tsukuba 305-0047, Japan}
\affiliation{$^{8}$Stavropoulos Center for Complex Quantum Matter, Department of Physics and Astronomy, University of Notre Dame, Notre Dame, IN 46556, USA}
\affiliation{$^{9}$Department of Advanced Materials Science, The University of Tokyo, Kashiwa 277-8561, Japan}
\affiliation{$^{10}$Tokyo College, The University of Tokyo, Bunkyo-ku, Tokyo 113-8656, Japan }
\affiliation{$^{\S}$Current address: Department of Physics, University of Science and Technology of China, Hefei 230026, China}
\affiliation{$^\dagger$Current address: Paul Scherrer Institut, Forschungsstrasse 111, 5232 Villigen PSI, Switzerland}

\maketitle
\newpage

\begin{center}
\Large{Abstract}
\end{center}
Skyrmion lattices (SkL) in centrosymmetric materials typically have a magnetic period on the nanometer-scale, so that the coupling between magnetic superstructures and the underlying crystal lattice cannot be neglected. Here, we reveal the commensurate locking of a SkL to the atomic lattice in Gd$_3$Ru$_4$Al$_{12}$ via high-resolution resonant elastic x-ray scattering (REXS). 
Weak easy-plane magnetic anisotropy, demonstrated here by a combination of  ferromagnetic resonance and REXS, penalizes placing a skyrmion core on a site of the atomic lattice. Under these conditions, a commensurate SkL, locked to the crystal lattice, is stable at finite temperatures -- but gives way to a competing incommensurate ground state upon cooling. We discuss the role of Umklapp-terms in the Hamiltonian for the formation of this lattice-locked state, its magnetic space group, the role of slight discommensurations, or (line) defects in the magnetic texture,  and contrast our findings with the case of SkLs in noncentrosymmetric material platforms.
\newpage

\begin{figure}[h]
\centering
\includegraphics[trim=3.cm 0.8cm 2.5cm 0.8cm,clip,width=0.95\textwidth]{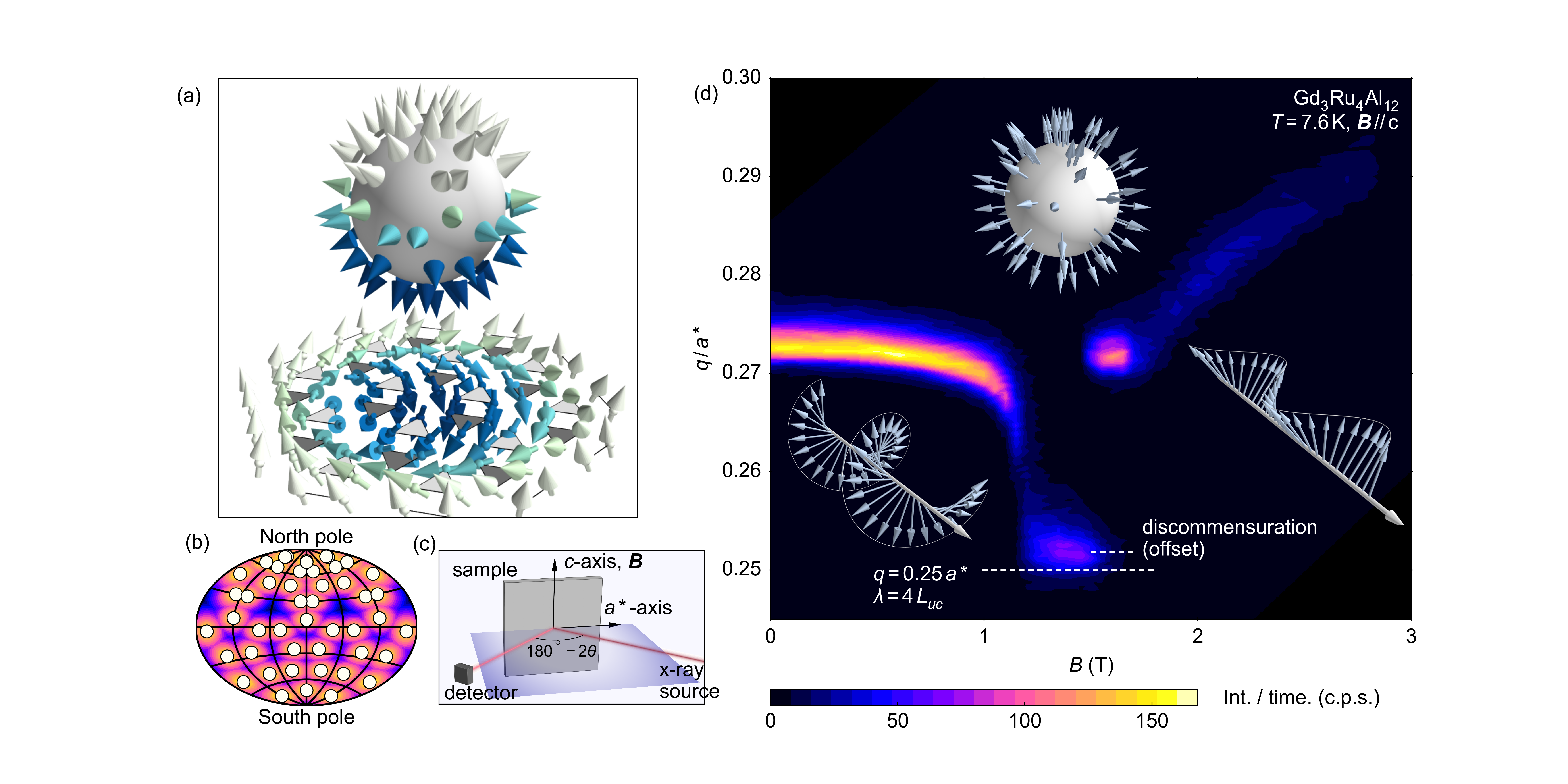}
\caption{Commensurate locking of the skyrmion lattice phase in Gd$_3$Ru$_4$Al$_{12}$ (GRA). (a) Magnetic moments of a skyrmion winding a sphere (top), and corresponding two-dimensional magnetic moment texture in real space (bottom). Each arrow corresponds to a site of the magnetic sublattice  (Supplementary Section \ref{Esec:spin_model}). (b) Map projection showing a hemisphere of (a) unfolded, with white dots indicating directions of magnetic moments. The colour code indicates the distance of each point on the sphere from the nearest moment on the sphere, in radians. (c) Experimental geometry of resonant elastic x-ray scattering (REXS) at the Gd-$L_2$ absorption edge, where the pink line is the trajectory of the x-ray beam. (d) Scattering intensity in REXS as a function of magnetic field ($B$) and momentum transfer $q$ from $\mathbf{q}= (q, 0, 0)$; $q= 0.25$ corresponds to $\lambda = 2\pi / q = 4\, L_{uc}$, four times the projection of the lattice constant parallel to $\mathbf{q}$ [Supplementary Fig.~\ref{Fig_S9}]. Incommensurate (IC) proper screw, commensurate skyrmion lattice (C-SkL), and IC fan phases are illustrated by insets. In C-SkL, the periodicity of the magnetic texture is locked to the crystal lattice up to a weak discommensuration (offset) of $\Delta q = 0.0018\,\,$r.l.u. 
$B$ denotes magnetic induction after demagnetization correction (Supplementary Fig. \ref{Fig_SI_phase_diagram}). Supplementary Figs. \ref{Fig_S_Bdep_7p6K} and \ref{Fig_S_Bscan_fitparams_7p6K} show raw data used to create this colormap and the results of Gaussian fitting to the data, respectively.}\label{Fig1}
\end{figure}

\begin{center}
\Large{Main text}
\end{center}
Magnetic skyrmion lattices (SkLs) are periodic arrays of vortex-like structures. In SkLs, magnetic moments are twisted into a knot, covering all directions of a sphere as we traverse a single magnetic unit cell (UC) [Fig. \ref{Fig1}(a)] \cite{Bogdanov89, Muehlbauer09, Tokura21}. These vortices were first described as topological solitons using the concepts of field theory, and such continuum models are most suitable when the magnetic UC is at least two orders of magnitude larger than the underlying crystallographic UC \cite{Binz06, Binz06b, Tokura21}. With a focus on frustrated, i.e. competing, interactions, recent theoretical work \cite{Okubo12, Hayami16, Ozawa17, Hayami17, Wang20} has proposed SkL formation in a high-symmetry context without spin-orbit driven Dzyaloshinskii-Moriya interactions, paving the way for the experimental observation of SkL phases with magnetic period on the nanometer-scale in centrosymmetric insulators and metals~\cite{Tokura21,Hirschberger19, Khanh20}. These quasi-discrete SkLs have raised hopes of enhanced functional responses, especially those related to the interplay of emergent electromagnetic fields with conduction electron (Bloch) waves, or with incident light waves \cite{Binz08, Heinze11, Nagaosa19, Sorn21, Tokura21, Kato23}. 

Evidence for coupling between the atomic lattice and skyrmion textures with lattice spacing $2-3\,$nanometers has emerged in tetragonal magnets: Centrosymmetric alloys host square and rhombic skyrmion lattices~\cite{Tokura21,Takagi22}, and non-centrosymmetric EuNiGe$_3$ exhibits a fascinating helicity reversal upon entering the SkL phase, where the magnetic texture breaks the sense of rotation prescribed by its polar structure~\cite{Matsumura23}. A key open challenge is the demonstration of a commensurate locking (C-locking) transition of the SkL's spin superstructure to the underlying lattice potential in such a centrosymmetric bulk material. This phenomenon is conceptually related to instabilities of the skyrmion vortex core anticipated for spin-$1$ systems~\cite{Zhang23} and its observation would provide a bridge between -- usually -- large-scale SkL spin textures and canted antiferromagnetism on the scale of a single unit cell. Indeed, theory predicts such C-locking based on Ruderman-Kittel-Kasuya-Yosida (RKKY) interactions and magnetic anisotropy, when the length scale of magnetic textures approaches the size of a crystallographic UC~\cite{Hayami21}.  Among inversion breaking material platforms, the hcp-Fe/Ir(111) interface has been reported to exhibit  C-locking using imaging techniques, although fcc-Fe/Ir(111) forms a lattice-incommensurate structure~\cite{Heinze11, Bergmann15, Wiesendanger16, Gutzeit23}. However, such locking between the periodicity of a magnetic skyrmion lattice and the underlying crystal structure, which can be expected when the length scale of magnetic textures approaches the size of a crystallographic UC~\cite{Hayami21}, has never been observed in a bulk material.

Using precise resonant x-ray measurements, we report a commensurate skyrmion lattice (C-SkL) in bulk samples of the centrosymmetric intermetallic Gd$_3$Ru$_4$Al$_{12}$, locked to the distorted kagome network of magnetic gadolinium ions, and surrounded by incommensurate (IC) phases.
We discuss this state based on (weak) single-ion anisotropy $K_1$, as supported by electron-spin resonance experiments.

For large classical spins, calculations on both triangular and breathing kagome lattices show that, if the single-ion anisotropy is of easy-plane type (easy-axis type), a commensurate skyrmion vortex may gain energy by locating its core at an interstitial site (on a lattice site)~\cite{Hayami21, SI}; an incommensurate skyrmion lattice does not benefit from this type of energy gain (Supplementary Fig. \ref{Fig_S8}). We illustrate this point in the lower part of Fig. \ref{Fig1}(a), depicting a realistic model for the C-SkL of Gd$_3$Ru$_4$Al$_{12}$ , described by normalized vectors $\mathbf{n}(x,y)$, which is mapped onto a sphere using a stereographic projection (Supplementary Section \ref{Esec:stereographic}). Here,  magnetic moments are conspicuously absent at the poles [Fig. \ref{Fig1}(a), upper]. The sparsity of magnetic moments at the poles becomes more apparent when unfolding the sphere using a cartographic projection (Fig. \ref{Fig1}(b) and Supplementary Section \ref{Esec:spin_model}). \\

\textbf{Observation of a commensurate skyrmion lattice (C-SkL)}\\
We use elastic x-ray scattering in resonance with the $L_{2,3}$ absorption edge of gadolinium, setting the sample in reflection geometry to precisely detect the magnetic period of each magnetic phase (Fig. \ref{Fig1}(c) and  Supplementary Section \ref{Esec:rexs_experimental}). Reporting data from a synchrotron light source, Fig. \ref{Fig1}(d) depicts the core observation of this work: At moderate temperatures $T = 7-8\,$K, a magnetic field $\mathbf{B}$ applied along the $c$–axis drives the incommensurate proper screw ground state (IC-PS) into a commensurate skyrmion lattice phase (C-SkL), and again to incommensurate fan-like order (IC-Fan).  The C-locking of the SkL was not observed in a previous study~\cite{Hirschberger19}, which included measurements on a $^{160}$Gd enriched single crystal with enhanced disorder (Supplementary Fig. \ref{Fig_SI_phase_diagram}). Let $a^*$ and $c^*$ be the reciprocal space lattice constants for our target material Gd$_3$Ru$_4$Al$_{12}$ in the hexagonal $P6_3/mmc$ space group, where magnetic gadolinium ions form a kagome (star of David) lattice with a breathing distortion, corresponding to alternating bond distances. In zero magnetic field, the magnetic modulation wavevector is $\mathbf{q} = (q, 0, 0)$ with $ q \approx 0.272\,a^*$ or wavelength $\lambda \approx 3.7\, L_{uc}$, where $L_{uc}$ is the dimension of the crystallographic UC projected parallel to $\mathbf{q}$ (Supplementary Fig. \ref{Fig_S9}). The wavevector's length $q$ decreases with $B$ in IC-PS, approaching a discontinuous jump at the first-order transition~\cite{Hirschberger19} towards $q = 0.25\,a^*$ in C-SkL. The magnetic period $\lambda$ in C-SkL is very close to $4\, L_{uc}$, with a slight offset indicated by two dashed horizontal lines in Fig. \ref{Fig1}(d). The role of this slight offset, or discommensuration $\lambda_\mathrm{disc}$, is discussed below. To further support the C-locking at $q = 0.25\,a^*$, we demonstrate in Supplementary Fig. \ref{Fig_S_Bscan_fitparams_7p6K} that the peak profiles are within less than one standard deviation from the commensurate value, and in Supplementary Fig. \ref{Fig_S_Tscan_fitparams_1p3T} that $q$ has weak temperature dependence in C-SkL as compared to IC-PS.

In the regime labeled as C-SkL in Fig. \ref{Fig1}, previous real-space imaging experiments have observed vortex structures~\cite{Hirschberger19}, and precise Hall effect measurements demonstrate that the noncoplanar magnetic state that is easily destroyed by a slight in-plane magnetic field~\cite{Hirschberger21b}. However, neither this prior work nor the present REXS  technique are able to determine whether the skyrmion core is located on a crystallographic lattice site, or on an interstitial site. This question of the relative phase shift between magnetic texture and crystal lattice can be addressed by measurements of single-ion anisotropy, combined with theoretical modeling: We turn to the  ferromagnetic resonance ( FMR) technique in the following section.\\

\begin{figure}[h]
\centering
\includegraphics[trim=0.5cm 0.1cm 1.5cm 0.8cm,clip,width=0.7\textwidth]{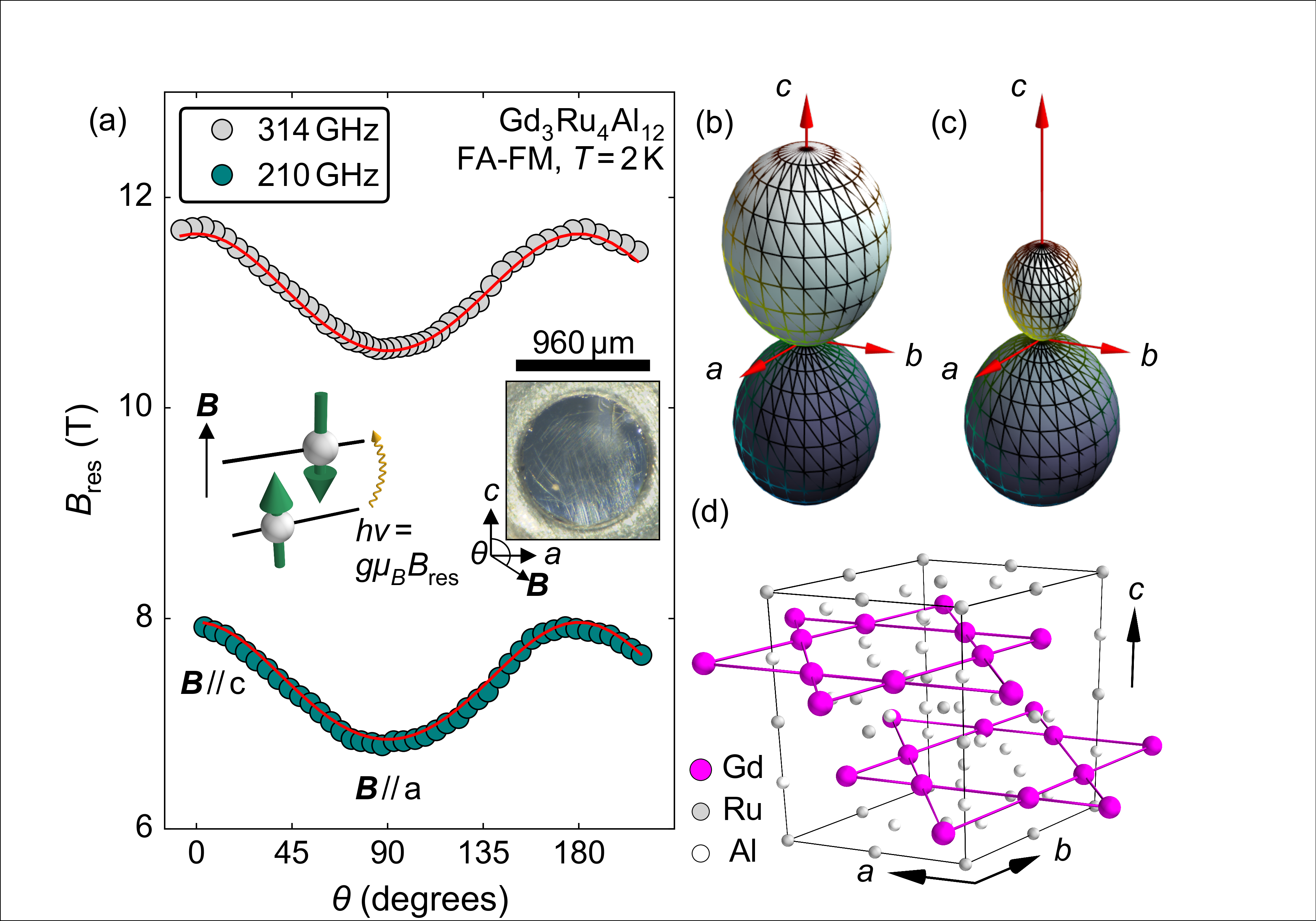}
\caption{Magnetic anisotropy in field-polarized ferromagnetic (FA-FM), incommensurate proper screw (IC-PS), and skyrmion lattice (C-SkL) phases of Gd$_3$Ru$_4$Al$_{12}$. (a) Resonance field $B_\mathrm{res} = \mu_0 H_\mathrm{res}$ in  ferromagnetic resonance (FMR) experiments in FA-FM at $210$ and $314\,$GHz (grey and green markers). Red lines: Fit to the data according to Smit-Beliers-Suhl, corresponding to weak in-plane anisotropy (Supplementary Section \ref{Esec:fmr_experimental}). Left inset: principle of  FMR, where transitions between up and down moments are induced by microwave radiation in a cavity. Right inset: geometry of magnetic field $\mathbf{B}$ with respect to the principal axes of a disk-shaped single crystal. (b,c) Anisotropic energy landscape $F_\mathrm{anis} = K_1 \cos^2(\theta) – E_Z \cos(\theta)$ in zero field (Zeeman term $E_Z = 0$) and finite magnetic field ($E_Z / K_1 = 0.3$), as a function of the direction of the bulk magnetization vector $\mathbf{M}$. (d) Crystal structure of Gd$_3$Ru$_4$Al$_{12}$ with magnetic breathing kagome sublattice of gadolinium, shown in magenta.}\label{Fig2}
\end{figure}

\textbf{Single-ion anisotropy in Gd$_3$Ru$_4$Al$_{12}$}\\
We prepared a cylindrical disk-shaped sample for  ferromagnetic resonance (FMR) experiments, spanned by the crystallographic $a$ and $c$ directions [Fig. \ref{Fig2}(a), right inset]. This highly symmetric geometry allows for simple data analysis when rotating the magnetic field in the plane of the disk (Supplementary Section \ref{Esec:fmr_experimental}). In the experiment, we drive the crystal into the field-aligned ferromagnetic (FA-FM) state with a large magnetic field, irradiate it with microwaves of frequency $\nu  = 210$ or $314\,$GHz, and observe a change of its reflectivity when the microwaves excite a resonance between moment-up and -down states [Fig. \ref{Fig2}(a), left inset]. 

In Fig. \ref{Fig2}(a), the anisotropic part of the free energy $F_\mathrm{anis}$ is deduced in FA-FM from a fit to the angular dependence of the  FMR resonance field $B_\mathrm{res}$; we disregard anisotropy constants beyond the first order (Supplementary Section \ref{Esec:fmr_experimental}). Using the saturation magnetization $M_S= 7\,\mu_B / \mathrm{Gd}^{3+}$, our fit yields easy-plane anisotropy with $K_1 = -0.13\,\mathrm{meV} / \mathrm{Gd}^{3+}$ for the potential $F_\mathrm{anis} = K_1 \cos^2(\theta)$. Therefore, the anisotropy field $B_\mathrm{ani}$ can be calculated as $\mu_0\left|2K_1\right|/M_S = 0.74\,$Tesla. As compared to magnetization measurements in Supplementary Fig. \ref{FIG_S_magnetization_anisotropy}, the present FMR study yields enhanced precision; field-dependent FMR experiments separate $g$-factor anisotropy and exchange anisotropy from the single-ion term (Supplementary Section \ref{Esec:fmr_experimental}). Figure \ref{Fig2}(b,c) illustrates the resulting iso-energy surfaces $F_\mathrm{anis}(\theta,\varphi)$ in zero (finite) magnetic field along the $c$-axis, respectively, where the spherical coordinates refer to the direction of the sample's bulk magnetization $\mathbf{M}$. \\

\begin{figure}[h]
\centering
\includegraphics[trim=1.3cm 0.1cm 1.cm 0.2cm,clip,width=0.85\textwidth]{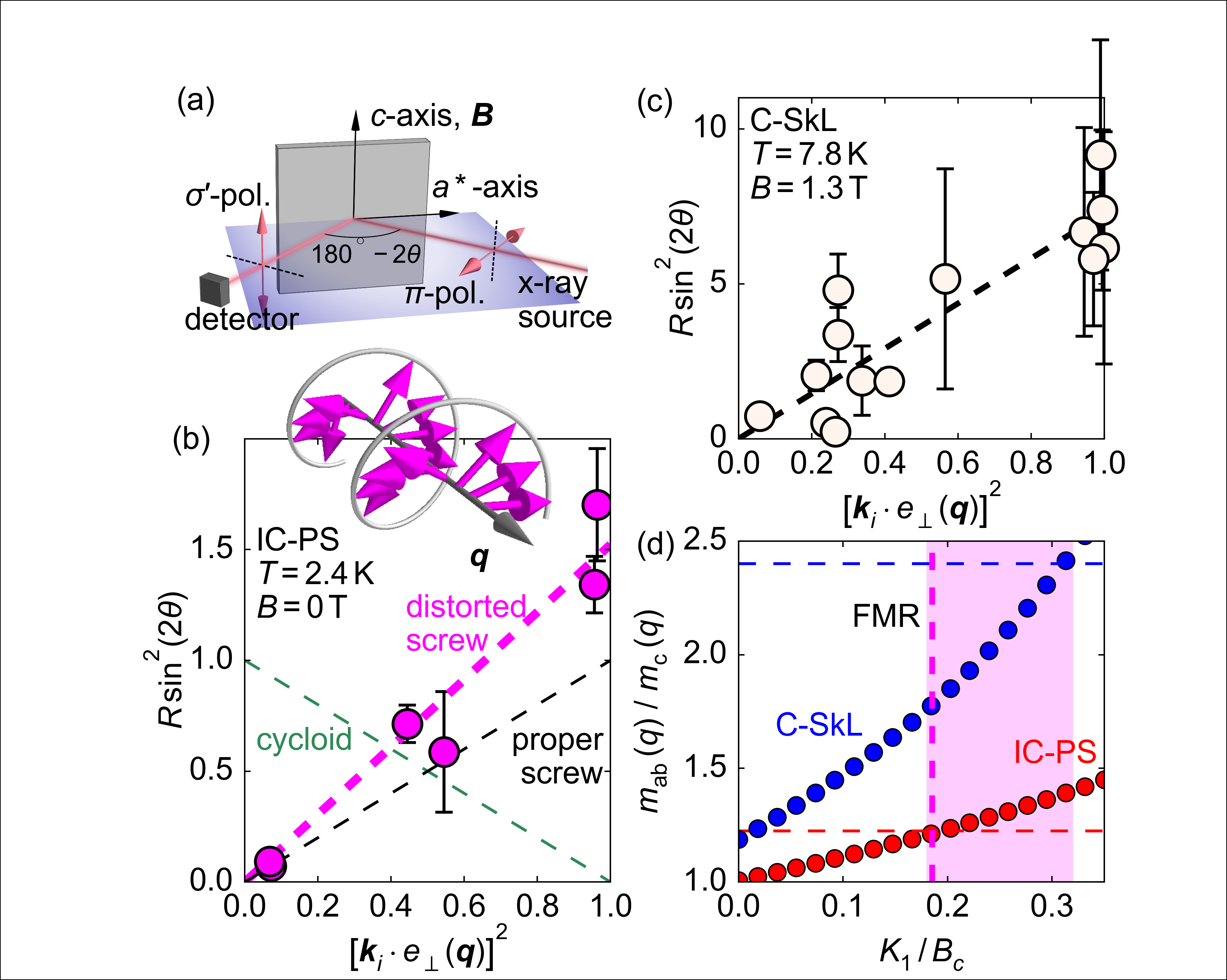}
\caption{Magnetic anisotropy in ordered states of Gd$_3$Ru$_4$Al$_{12}$. (a) Experimental geometry of resonant elastic x-ray scattering (REXS) with [in contrast to Fig. \ref{Fig1}(c)] polarization analysis. $\pi$ and $\sigma$ indicate x-rays (photons) polarized parallel and perpendicular to the scattering plane (purple), respectively. (b) Elliptic distortion of IC-PS determined by polarization analysis in REXS. $I_{\pi-\sigma^\prime}$, $I_{\pi-\pi^\prime}$, $\mathbf{k}_i$, and $\mathbf{e}_\perp$ are integrated intensities in the polarization-flipped and polarization-conserved scattering channels, the incoming beam's wavevector, and the in-plane direction perpendicular to the propagation direction $\mathbf{q}$ of IC-PS, as in Eq. (\ref{Eq1}). Green, black, and magenta dashed lines indicate model predictions for spherical cycloid, spherical proper screw, and elliptically distorted screw, the latter being shown in the inset cartoon. The value of $R\sin^2(2\theta)$ at the right boundary of the panel corresponds to the elliptic distortion $(m_\mathrm{ab}/m_c)^2$ as defined in Eq. (\ref{Eq1}). (c) Analogous for the commensurate skyrmion (C-SkL) state; see Table \ref{Etable:CSkL_intensity} for more details. Note the different $y$-axis range as compared to (b). (d) Calculated elliptic distortion of IC-PS and C-SkL as a function of anisotropy constant $K_1$, normalized to the critical field $B_c=4\,$T for transition to the field-aligned ferromagnetic (FA-FM) state. Dashed horizontal lines, dashed vertical line, and highlighted area indicate the experimental values for the ellipticity, the value of $K_1$ from  FMR, and and the range of $K_1$ consistent with REXS experiments, respectively.}\label{Fig3}
\end{figure}

\textbf{Anisotropy and anharmonic distortion of proper screw and skyrmion phases}\\
Easy-plane anisotropy ($K_1 < 0$) can also be verified semi-quantitatively by resonant elastic x-ray scattering in the ordered phases with periodic long-range order. Figure \ref{Fig1}(c) shows the geometry of our experiment with polarization analysis: the purple scattering plane is spanned by the wavevectors $\mathbf{k}_i$ and $\mathbf{k}_f$ of the incoming and outgoing x-ray beams, with polarization $\pmb{\varepsilon}_i$ and $\pmb{\varepsilon}_f$, respectively. We choose the incoming beam polarization $\pmb{\varepsilon}_i$ to lie within the scattering plane for all our experiments ($\pi$-polarization). While the data in Fig. \ref{Fig1}(d) represents a sum of scattered photons with all possible $\pmb{\varepsilon}_f$, we now add an analyser plate before the x-ray detector to separate two components of the scattered beam: $I_{\pi-\pi^\prime}$ and $I_{\pi-\sigma^\prime}$ with $\pmb{\varepsilon}_f$ within or perpendicular to the scattering plane, respectively. From the scattered intensities at various magnetic reflections, we extract the ratio (Supplementary Section \ref{Esec:rexs_experimental})
\begin{equation}
\label{Eq1}
R \sin^2(2\theta) = I_{\pi-\sigma^\prime} \sin^2(2\theta) / I_{\pi-\pi^\prime} =\left[\mathbf{k}_i \cdot \mathbf{m}_\mathrm{ab}(\mathbf{q}) / m_\mathrm{c}(\mathbf{q})\right]^2 
\end{equation}
with $\mathbf{m}_\mathrm{ab}$ the component of the modulated magnetic moment in the scattering plane. The expected behavior of IC-PS (spin plane $a$-$c$) and IC cycloid (spin plane $a^*$-$c$) is indicated in Fig.  \ref{Fig3}(b) by black and green dashed lines, respectively; a line with positive slope indicates  IC-PS character. Supplementary Fig. \ref{Fig_SI_polarization_analysis} shows representative raw data, as used to create this panel.

Beyond identifying the PS character of the spin modulation, this analysis allows us to estimate the effect of single-ion anisotropy in IC-PS. Specifically, we observe an elliptic distortion, i.e. a deviation from a simple harmonic screw model. Figure  \ref{Fig3}(b,c) displays maxima of $R \sin^2(2\theta)$ around $1.5\,$(IC-PS) and $7\,$(C-SkL), so that Eq. (\ref{Eq1}) delivers $m_\mathrm{ab}(\mathbf{q}) / m_\mathrm{c}(\mathbf{q})$ around $1.2\,$(IC-PS) and $2.6\,$(C-SkL). Figure \ref{Fig3}(b), inset, illustrates the proposed anharmonic distortion in IC-PS, where magnetic moments prefer to tilt towards the basal plane  to gain anisotropy energy. In Fig. \ref{Fig3}(d), a simulation of REXS anisotropy as a function of $K_1$ for a spin model of both IC-PS and C-SkL(Supplementary Section \ref{Esec:spin_model}) shows that  the experimentally observed ellipticity is consistent with the results of  FMR in Fig. \ref{Fig2}.  Given the robust observation of  $K_1\,$$<0$ via various experimental techniques, we conclude that a C-SkL in magnetic space group $P6_32^\prime 2^\prime$ and 
with skyrmion core \textit{between lattice sites} is stable in Gd$_3$Ru$_4$Al$_{12}$ (Supplementary Section \ref{Esec:magnetic_space_group}). \\

\begin{figure}[h]%
\centering
\includegraphics[trim=2.8cm 0.2cm 1.8cm 0.2cm,clip,width=1.\textwidth]{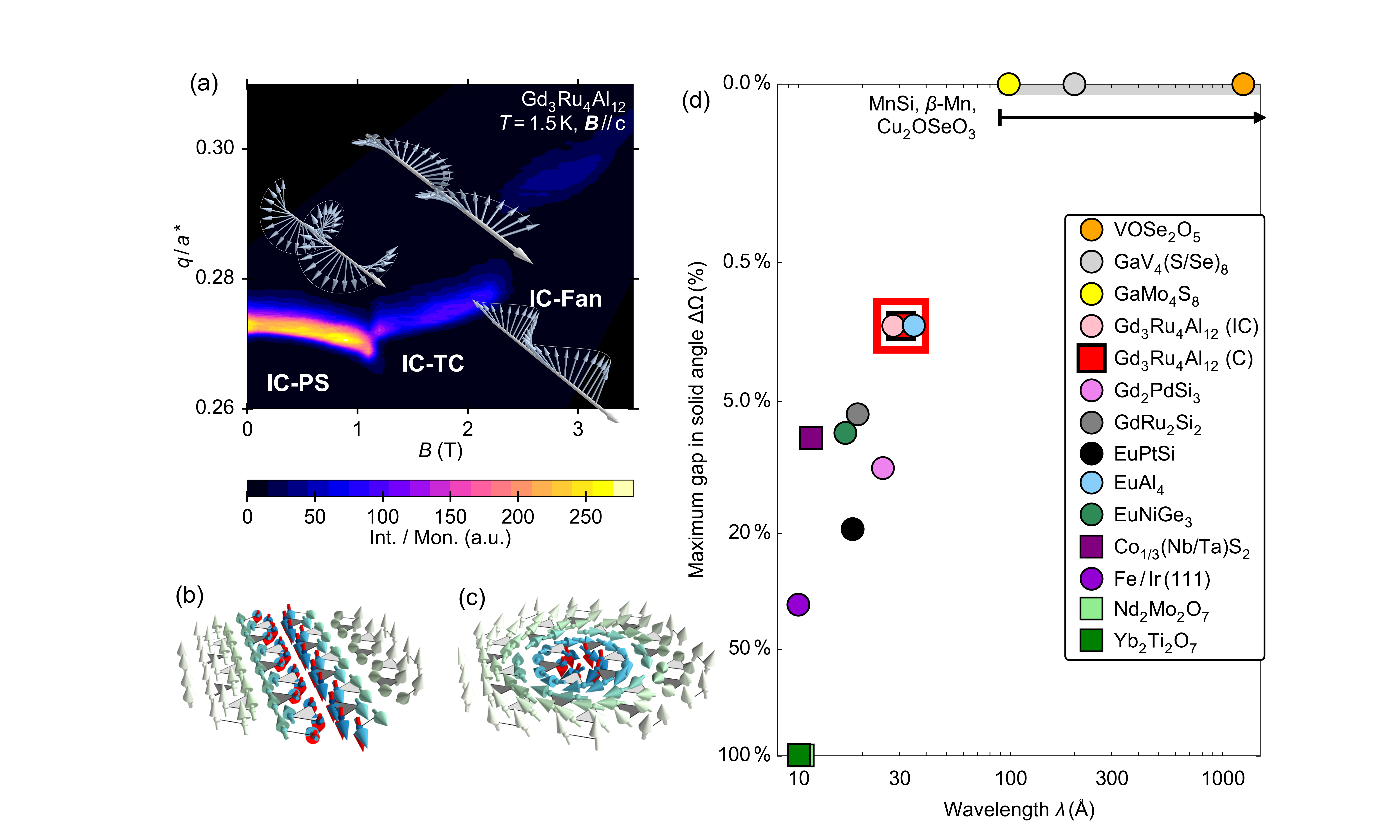}
\caption{Stability of the commensurate skyrmion lattice (C-SkL) as compared to other magnetic phases. (a) Resonant x-ray scattering intensity, normalized to monitor counts, at base temperature ($T=1.5\,$K). Three incommensurate (IC) magnetic phases are shown: from left to right, proper screw (IC-PS), transverse conical (IC-TC), and fan-like (IC-Fan) as illustrated by insets. Quantitative analysis of Gaussian line-scan profiles is given in Supplementary Fig. \ref{Fig_S_Bscan_fitparams_1p5K}. (b,c) Distortion of proper screw (left) and skyrmion textures in anisotropy potential and external magnetic field along the $z$-direction (Supplementary Section \ref{Esec:spin_model}). The undistorted moment directions (red) are superimposed on those moments that have been rotated by more than a critical angle. (d) Relationship of magnetic texture dimensions ($\lambda$) and coverage of directions on the sphere for various materials with noncoplanar textures and spin chirality~\cite{Gardner10,Hirschberger19,Tokura21,Takagi23,Matsumura23, Butykai22}. At the center of the plot, the commensurate C-SkL state in Gd$_3$Ru$_4$Al$_{12}$ is highlighted by a red box. On the $y$-axis, a continuous magnetic texture has zero \textit{uncovered solid angle}, i.e. we assign a value of $0\,\%$.}\label{Fig4}
\end{figure}

\textbf{Competition of commensurate and incommensurate phases}\\
Figure \ref{Fig4}(a) shows a magnetic field scan of x-ray scattering intensity at the lowest accessible temperature, $T = 1.5\,$K; see Supplementary Fig. \ref{Fig_SI_phase_diagram} for detailed phase diagrams and comparison to prior work. In contrast to Fig. \ref{Fig1}(d), the C-SkL phase is now absent, being replaced by an incommensurate transverse conical (IC-TC) state~\cite{Hirschberger19}. 

We consider the observation of C-locking induced by magnetic fields, in our experiment, based on the single-ion contribution to the spin Hamiltonian (Supplementary Section \ref{Esec:Fourier_modes})
\begin{equation}
\mathcal{H}_\mathrm{anis} \propto -K_1 \sum_{\mathbf{G}}\sum_{\mathbf{q}\in\mathrm{BZ}} \mathcal{S}^z(\mathbf{q})\mathcal{S}^z(\mathbf{G}-\mathbf{q})\Big[\sum_{\mathbf{d}\in\mathrm{u.c.}} e^{i\mathbf{G}\cdot\mathbf{d}} \Big]
\label{eq:hamiltonian}
\end{equation}
where $\mathbf{r} = \mathbf{R} + \mathbf{d}$ is the position of a magnetic ion, decomposed into a unit cell coordinate and an intra-cell coordinate; $\mathbf{q}$ and $\mathbf{G}$ are the momentum in the first Brillouin zone and a reciprocal lattice vector, respectively. Typically, a small number of Fourier modes $\mathbf{q} = \mathbf{q}_\nu$ and $\mathbf{G}=0$ are selected in models of incommensurate helimagnetic ordering~\cite{Khanh20,Hirschberger21b,Takagi22}. Meanwhile, C-locking is favored by the Umklapp terms $\mathbf{G}\neq 0$~\cite{Wilson1975, Fleming1980}, representing a coupling between the primary Fourier mode and its higher harmonics in helimagnets. Specifically for the $q = 0.25\,a^*$ C-SkL in Gd$_3$Ru$_4$Al$_{12}$, $\pmb{\mathcal{S}}(3\mathbf{q})\cdot \pmb{\mathcal{S}}(\mathbf{q})$ and $\pmb{\mathcal{S}}(2\mathbf{q})^2$ are the leading contributors, adding to $\mathbf{G} = (1, 0, 0)$ and coupling the helimagnetic order to the lattice. Application of a magnetic field to IC-PS enhances the elliptic distortion as demonstrated in Fig. \ref{Fig3}, amplifies anharmonicity, shifts $\mathbf{q}$ away from the value preferred by the exchange interaction and towards commensurability, and ultimately induces C-locking between the spin texture and the underlying lattice. In Supplementary Section \ref{Esec:Fourier_modes}, we thus derive an expression for the energetic contribution that depends on the position of the skyrmion core. Nevertheless, a full numerical treatment of Eq. (\ref{eq:hamiltonian}) with large numbers of Fourier modes, to capture changes in the optimal $\mathbf{q}$ -- as well as changes in the spin texture -- as a function of magnetic field and temperature, remains a challenge at present (Supplementary Section \ref{Esec:spin_model}).

More intuitively,  we initialize C-SkL and C-PS on the lattice and allow the magnetic moments to relax in the combined potential of exchange interaction, magnetic anisotropy, and external field of material specific strength [white / green arrows in Fig. \ref{Fig4}(b,c)]. The red arrows show a significant distortion of the textures especially around the south pole. The density of strongly distorted moments is much larger for the C-PS state with quasi one-dimensional spin texture, which has an  extended south pole region. In particular, the south pole direction of the magnetic moment sphere in Fig. \ref{Fig1}(a) corresponds to a point (to a line) in case of a SkL (of a PS). We also calculated the $z$-projection of magnetic moments for a spin model of C-PS, C-SkL, and other orders, using a reasonable anisotropy value of $K_1$ based on FMR, numerically confirming the favorable pinning of the C-SkL due to stronger higher harmonics (Supplementary Fig. \ref{Fig_S8}).\\

\textbf{Discussion and Outlook}\\
Figure \ref{Fig1}(d) reveals a discommensuration of the magnetic lattice, i.e. a slight offset from $q/a^*=0.25$ that indicates occasional defects. The effect is well understood in one-dimensional chain systems~\cite{Bak82}: For example, the introduction of small amounts of chemical disorder causes proliferation of discommensurations in spin-Peierls chains, as manifested in a drop of $\lambda_\mathrm{disc}$ that ultimately destroys the commensurate (C) order~\cite{Hase93, Kiryukhin95a, Kiryukhin95b, Kiryukhin96}. Discommensurations in two dimensions, mostly line defects, also appear for surface-adsorbed atoms and in the multi-directional charge-density wave state of transition metal dichalcogenides such as $2H$-TaSe$_2$, where C-IC transitions have been extensively studied~\cite{Wilson1975, Fleming1980}. Among materials with two-dimensional spin textures, we compare to the C-SkL observed in an interfacial system~\cite{Bergmann15,Wiesendanger16,Gutzeit23}: For inversion-breaking hcp-Fe/Ir(111) (for centrosymmetric Gd$_3$Ru$_4$Al$_{12}$), the spin structure is found to be commensurate within less than $10\,\%$ (within $0.7\,\%$) of the magnetic period, and there is no (there is) evidence of C-IC transitions by cooling or application of a magnetic field. Based on measurements of magnetic anisotropy and modeling, we argue here that Gd$_3$Ru$_4$Al$_{12}$ has skyrmion cores on interstitial lattice sites, as does hcp-Fe/Ir(111); further that the discommensuration represents the appearance of line defects (characteristic spacing $\lambda_\mathrm{disc} = 430\,$nm) in the former -- while phase-slip domain walls (spacing $\sim10\,$nm) and domains of the net magnetization (spacing $\sim30\,$nm) have been observed in the latter. We note that domain walls of the skyrmion helicity may appear in the bulk C-SkL of centrosymmetric materials~\cite{Okubo12,Leonov15}; these are forbidden in inversion breaking platforms such as hcp-Fe/Ir(111).

Next, we  compare the experimental and theoretical results for Gd$_3$Ru$_4$Al$_{12}$ to earlier studies of magnetism in elemental rare earth metals. Various scenarios for the interplay of commensurate (C) and IC phases have been advanced~\cite{Koehler72,Jensen91}: One typical observation is the appearance of spin-slip structures in the IC order, and eventual \textit{squaring up} at low temperatures. In this scenario, thermal fluctuations at higher $T$ locally reduce the ordered magnetic moment, allowing for the formation of IC magnetism close to the N{\'e}el point, e.g. in elemental Holmium ~\cite{Koehler72}.  Secondly, when single-ion anisotropy is much weaker than the dominant exchange interactions, C orders appear at elevated temperatures before the system settles into an IC ground state (e.g. Ref. \cite{Venturini96}).  The present study generalizes the second scenario to the case of complex, twisted magnetic textures, such as the C-SkL. 

Finally, Fig. \ref{Fig4}(d) summarizes recent progress on material search and magnetic structure studies of noncoplanar magnets with net scalar spin chirality, especially SkL host compounds~\cite{Gardner10,Tokura21,Matsumura23,Takagi23}. As the dimension $\lambda$ of the magnetic UC increases, magnetic moments cover all directions of the $S_2$ sphere, and the maximum \textit{gap in angular coverage} (as defined in percent of $4\pi$) approaches zero. For example, noncoplanar antiferromagnets on the left side of Fig. \ref{Fig4}(d) leave large fractions of $S_2$ remain out of reach of magnetic moments. Being located at the center of the plot, the C-SkL in the centrosymmetric, breathing kagome magnet Gd$_3$Ru$_4$Al$_{12}$ represents an essential link between long-period, incommensurate magnetic textures, stabilized e.g. by Dzyaloshinskii-Moriya interactions in MnSi~\cite{Bak80,Tokura21}, and so-called topological antiferromagnets with tilted magnetic moments, such as pyrochlore systems~\cite{Gardner10}. Although the present C-SkL is locked to the crystal lattice due to weakly in-plane magnetic anisotropy, as confirmed here by ferromagnetic resonance experiments, the magnetic unit cell is large enough that moments densely cover $S_2$, leaving only marginal gaps on the order of $5\,\%$ of $4\pi$.



%


\newpage

\newpage

\textbf{Acknowledgements}\\
We thank Taro Nakajima for support regarding resonant elastic x-ray scattering (REXS) experiments at beamline BL-3A of Photon Factory. Akiko Kikkawa provided guidance regarding single crystal synthesis. We acknowledge Ferenc Simon for fruitful discussions and technical support at EPFL, and Dieter Ehlers for providing his software to model the ferromagnetic resonance data. We are also grateful to Vladimir Tsurkan for polishing the samples for FMR. REXS measurements at the Institute of Material Structure Science of the High Energy Accelerator Research Organization (KEK) and at SPring-8 BL19LXU
 were carried out under the approval of the Photon Factory program advisory committee (Proposal No. 2020G665) and under grant number 20210007, 
 respectively. M.He. and H.-A. K.v.N. acknowledge funding within the joint RFBR-DFG research project contract No. 19-51-45001 and KR2254/3-1. 
 Ma.Hi. benefited from JSPS KAKENHI Grant No. JP22H04463 and 23H05431, while also acknowledging grants by the Murata Science Foundation, the Yamada Science Foundation, the Hattori Hokokai Foundation, the Iketani Science and Technology Foundation, the Mazda Foundation, the Casio Science Promotion Foundation, the Inamori Foundation, and the Kenjiro Takayanagi Foundation. This work is partially funded by the Deutsche Forschungsgemeinschaft (DFG, German Research Foundation) under project numbers 518238332 and TRR 360–492547816 and by the Japan Science and Technology Agency via Core Research for Evolutional Science and Technology (CREST) Grant Nos. JPMJCR1874, JPMJCR20T1 (Japan), and by FOREST No. JPMJFR2238.\\

\clearpage
\begin{center}
\Large{Main text Figures}
\end{center}

\clearpage

\renewcommand\thefigure{S\arabic{figure}} 
\renewcommand\thesection{S\arabic{section}} 
\renewcommand\thetable{S\arabic{table}} 
\setcounter{figure}{0} 
\setcounter{section}{0}  
\setcounter{table}{0} 

\begin{center}
\Large{Supplementary Information}
\end{center}

\section{Model parameters for skyrmion texture}
\label{Esec:spin_model}
In the hexagonal $P6_3/mmc$ structure of Gd$_3$Ru$_4$Al$_{12}$, the angle between the crystal lattice directions $\mathbf{e}_a$ and $\mathbf{e}_b$ is $120^\circ$. These directions, or unit vectors, are termed $a$, $b$, and $c$ axes in the main text, and the corresponding lattice constants are $a$, $b = a$, and $c$. Meanwhile, the primary directions of reciprocal (momentum) space are $\mathbf{e}_{a^*}$ ($\mathbf{e}_{b^*}$) at a $30^\circ$ ($90^\circ$) angle with respect to $\mathbf{e}_a$, as shown in Supplementary Fig. \ref{Fig_S9}(a); $\mathbf{e}_{c^*}$ is parallel to $\mathbf{e}_{c}$. These axes (unit vectors) are termed $a^*$ axis and so on in the main text, and the corresponding lattice constants (in $\mathbf{q}$-space) are $a^* = b^* = 4\pi / (\sqrt{3} a)$ and $c^* = 2\pi / c$.

Figure \ref{Fig1}(a) shows an approximated, atomistic magnetic texture for the commensurate skyrmion lattice on the breathing kagome lattice of Gd$_3$Ru$_4$Al$_{12}$, described by a harmonic superposition of three screws with modulation vector $\mathbf{q}_1 = 0.25\,a^*\, \mathbf{e}_{a^*}$ and two more, equivalent directions:
\begin{align}
\mathbf{M}_i(\mathbf{r}) &= \mathbf{e}_z\cos\left(\mathbf{q}_i\cdot\mathbf{r}+\phi_i\right)+\mathbf{e}_{\perp i} \sin\left(\mathbf{q}_i\cdot\mathbf{r}+\phi_i\right) \label{eq:mag_spiral}\\
\mathbf{M}(\mathbf{r}) &= \mathbf{M}_1(\mathbf{r}) +\mathbf{M}_2(\mathbf{r})+\mathbf{M}_3(\mathbf{r})
\label{eq:mag_skyrmion}
\end{align}
where $\mathbf{e}_z$ ($\mathbf{e}_{\perp i}$) denotes the crystallographic $c$ axis (the direction perpendicular to both $z$ and $\mathbf{q}_i$). The phases $\phi_i$ obey $\Sigma_i \phi_i = 0$ ($\mathrm{mod}\, \pi$) and, in the discussion of the main text, are adjusted so as to place the core of the skyrmion on an interstitial site of the crystal lattice.  In Fig. \ref{Fig1}, Gd$^{3+}$ sites are depicted as coloured spheres with arrows representing the local magnetic moment $\mathbf{m}$, the $z$-component of which corresponds to the colour scale. Figure \ref{Fig2}(d) shows the upper and lower rare earth layers of Gd$_3$Ru$_4$Al$_{12}$ in each UC; these are collapsed into a single plane in the illustrations of Figs. \ref{Fig1}, Supplementary Fig. \ref{Fig_S9}.
Small triangles of the distorted kagome structure are highlighted in dark (light) grey for the upper (lower) layer in Fig. \ref{Fig1}.

Figure \ref{Fig4}(b,c) shows distorted magnetic structures calculated as follows. A harmonic proper screw, or a superposition of three harmonic screws (skyrmion) is initialized on the lattice according to Eq. (\ref{eq:mag_skyrmion}). The length of all magnetic moments is normalized. An energy functional is minimized which describes the balance of an exchange field $B_{ex}=1$ -- parallel to the undistorted direction of each moment --, an anisotropic potential $K_1 \cos^2(\theta)$, and a magnetic field term $-B\cos(\theta)$. In the figure, we choose $K_1 = 0.2$ and $B = 0.3$ for moderately strong distortion. Moments whose polar angle $\theta$ is distorted by more than $36^\circ$ are depicted in red.

\section{Magnetic space group of commensurate skyrmion lattice}
\label{Esec:magnetic_space_group}
In full, the space group of $R_3$Ru$_4$Al$_{12}$ ($R =$ rare earth) is represented as $P6_3/m\,2/m\,2/c$. While the coexistence of $\mathbf{q} = (0.25, 0, 0)$, $(0, 0.25, 0)$, and $(0.25, -0.25, 0)$ ordering vectors expands the $a$ and $b$ axes by four times each, the shape of the unit cell (UC) is kept unchanged. Furthermore, the presently observed Bloch skyrmions break all mirror and glide planes; such symmetries would reverse the vector spin chirality of the texture. Thirdly, the magnetic order's net magnetization along the $c$ axis breaks time reversal symmetry, with all two-fold rotation axes perpendicular to $c$ changed to $2^\prime$ axes: this notation indicates two-fold rotation followed by time reversal. The maximum symmetry of the present commensurate skyrmion state is therefore $P6_3\,2^\prime\,2^\prime$.

We further distinguish three cases, depending on the location of a skyrmion's core on the lattice. First, if the skyrmion core is coincident with the sixfold screw axis of Gd$_3$Ru$_4$Al$_{12}$, the screw axis survives. The magnetic space group is $P6_3\,2^\prime\,2^\prime$, and this is the state favored by easy-plane anisotropy as discussed in the main text. Second, if it is coincident with a threefold axis, i.e. with the center of a gadolinium triangle, the global screw symmetry is lowered to threefold. The magnetic space group is $P32^\prime$. 
Note that the skyrmion core is not coincident with the twofold axis along $[110]$. Thirdly, if the core is on a rare earth site, no rotation symmetry about the $c$ axis survives; each rare earth ion has $mm2$ site symmetry, with the $C_2$ rotation axis pointing into / out of a Gd$^{3+}$ triangle and aligned in the $a$-$b$ plane. The magnetic space group is $P 2^\prime$.

In the second and third cases, the $2^\prime$ axis is located on the kagome plane formed by gadolinium ions, and it is worth pointing out that there are two kinds of rare earth planes. Also note that in the discussion of magnetic space groups, we assumed absence of helicity domains, i.e. assumed that all skyrmions in a given crystal have the same, left- or right-handed (clockwise or counter-clockwise), helicity.

\section{Cartographic projection}
\label{Esec:stereographic}
For visualizing the magnetic texture in Fig. \ref{Fig1}(a), we mapped a magnetic skyrmion onto a sphere in three dimension, and subsequently mapped from the sphere onto a planar surface using the cartographic Nicolosi globular projection. Let ($\theta$, $\varphi$) denote a point where a line between ($x$, $y$) and the north pole of the sphere penetrates the sphere's surface. These spherical coordinates are identified with ($x$, $y$) in Fig. \ref{Fig1}(a), upper side, while a point on the sphere is projected back onto the plane using the formalism in Ref. \cite{Snyder89}.\\

Starting from polar angles $\theta$, $\varphi$ on the sphere, the $x$ and $y$ coordinates of the planar projection are defined as
\begin{align}
x &= \frac{\pi}{2}R\left(M + \mathrm{sgn}(\varphi-\varphi_0)\sqrt{M^2+\frac{\cos^2(\theta)}{1+b^2/d^2}}\right)\\
y &= \frac{\pi}{2}R\left(N - \mathrm{sgn}(\theta)\sqrt{N^2-\frac{d^2 \sin^2(\theta)/b^2+d\sin(\theta)-1}{1+d^2/b^2}}\right)
\end{align}
with functions
\begin{align}
M &= \frac{b \sin(\theta)/d - b/2}{1+b^2/d^2},\\
N &= \frac{d^2 \sin(\theta) / b^2 + d/2 }{1+d^2/b^2},\\
b &= \frac{\pi}{2(\varphi-\varphi_0)} - \frac{2}{\pi}(\varphi-\varphi_0),\\
c &= \frac{2}{\pi} \theta,\,\,\mathrm{and} \\
d &= \frac{1-c^2}{\sin(\theta)-c}.
\end{align}
The sphere's radius $R$ and the reference point for the longitude $\varphi_0$ can be set to $1$ and $0$, respectively.

\section{Sample preparation and characterization}
We prepared polycrstals of Gd$_3$Ru$_4$Al$_{12}$ 
by arc melting of the constituent elements in argon atmosphere, carefully turning pellets at least three times. 
Subsequently, single crystals were grown from these polycrystals by the floating zone technique under argon flow. Before the melting step, the halogen-lamp based furnace was evacuated to a base pressure of $8\cdot 10^{-4}\,$Pa, pumping for about three hours. The growth speed in the float zoning step was $2-4\,$mm/hr. We crushed single crystalline pieces into a fine powder for x-ray diffraction 
and refined the data by the Rietveld method using the RIETAN software package \cite{Izumi07}. Hence, impurity phases with volume fraction larger than $4\,\%$ were excluded by this analysis. Further, single-crystalline pieces were oriented using a Laue camera, cut with a diamond saw, and hand-polished to a high sheen ($\sim 1\,\mu$m grit) for single crystal x-ray diffraction measurements. Using a microscope with Nicolet prism, single-crystalline pieces with small amounts of RuAl$_3$ impurity phase, which forms tear-drop shaped inclusions of $< 2\,\%$ volume fraction that are hard to detect by laboratory x-rays, were excluded based on patterns in the surface-reflected light. Finally, magnetization measurements ($M$-$H$ curve at $T = 2\,$K, $M$-$T$ curve at $\mu_0H = 0.1\,$T) were used to confirm systematic evolution of long-range order in these single crystals.

\section{Ferromagnetic resonance measurements}
\label{Esec:fmr_experimental}
In order to avoid unwanted shape anisotropy effects, we polished the Gd$_3$Ru$_4$Al$_{12}$ samples cut within the $a$-$c$ plane into a cylindrical disc form, with ellipticity below $4\,\%$. The ratio of diameter to thickness was $12$ for the final disc, with a thickness of $120\,\mu$m. 

High-field  ferromagnetic resonance (FMR) measurements were performed at {\'E}cole Polytechnique F{\'e}d{\'e}rale de Lausanne (EPFL, Switzerland) using a home built, high-sensitivity, quasi-optical spectrometer operating in the range of $50-420\,$GHz. This instrument covers a broad magnetic field regime up to $16\,$T, using a superconducting magnet. Its variable temperature insert operates across the temperature range $1.5–300\,$K, using the dynamic flow of helium gas, or liquid helium, through a heat exchanger right below the sample space. For more details, see Ref. \cite{Nafradi08}.
   
The overall temperature accuracy of the system is $0.1\,$K. The polished disc-like sample is mounted on a goniometer with the $a$-$c$ plane coinciding with plane of rotation for the static magnetic field, the angular position of which is controllable and detectable via a potentiometer. Rotation proceeded in $5-10$ degree steps, at sample temperature $T = 2\,$K. The signal-to-noise ratio of the spectra is improved by recording the field-derivative of the absorbed power $dP/dH$ using a lock-in technique with magnetic field modulation. The angular-dependent resonance data can be evaluated using the Smit-Beliers-Suhl formula as described in Ref. \cite{Ehlers17}. Using the saturation magnetization $M_S = 7\,\mu_B / \mathrm{Gd}^{3+}$, our fit yields the easy-plane anisotropy constant $K_1 = -1.944\cdot 10^6\,$erg/cm$^3$ ($-0.13\,\mathrm{meV}\, /\, \mathrm{Gd}^{3+}$) and the temperature independent, isotropic $g$-factor $2.005$.\\

\section{Resonant x-ray scattering experiments (\texorpdfstring{G\MakeLowercase{d}}{Gd}-$L_{2,3}$)}
\label{Esec:rexs_experimental}
Resonant elastic x-ray scattering (REXS) experiments are carried out in reflection geometry at RIKEN beamline BL19LXU of SPring-8 and beamline BL-3A of Photon Factory, KEK, with the sample mounted inside a cryomagnet. The preparation of Sample A, used to obtain the data in Fig. \ref{Fig3}(b), is discussed in Ref.~\cite{Hirschberger19}. Sample B, which was used to obtain all other data in this manuscript, has a surface perpendicular to the $[110]$ axis, and was polished to reduce loss of intensity by diffuse scattering of x-rays. The energy of incident x-rays is matched to the $L_2$ or $L_3$ absorption edge of Gadolinium, where magnetic scattering involves virtual excitations from the $2p$ to the $5d$ atomic shells (Supplementary Fig. \ref{Fig_S1}). The $5d$ shell is coupled to the dominant magnetic species, the half-filled $4f$ orbitals, by intra-atomic exchange correlations.

For data in Figs. \ref{Fig1} and \ref{Fig4}, which are collected at the Gd-$L_3$ edge ($E_\mathrm{x-ray} = 7.243\,$keV) at BL19LXU of SPring-8, we do not carry out polarization analysis of the diffracted beam. For data in Figs. \ref{Fig3} and Supplementary Fig. \ref{Fig_SI_polarization_analysis}, which were collected at the Gd-$L_2$ edge ($E_\mathrm{x-ray} = 7.932\,$keV) at BL-3A of Photon Factory, $\pi-\sigma^\prime$ and $\pi-\pi^\prime$ components of the diffracted beam are separated using an analyser plate made from pyrolytic graphite (PG-006).

The details of the polarization analysis in the latter case are as follows. Let $\mathbf{k}_i$ ($\mathbf{k}_f$) be the wavevector of the incoming (outgoing) x-ray beam with polarization vector $\pmb{\varepsilon}_i$ ($\pmb{\varepsilon}_f$), where the scattering plane is spanned by the two wavevectors; c.f. purple plane in Fig. \ref{Fig1}(c). Let $z$ be the direction perpendicular to the scattering plane. The incident beam is $\pi$-polarized, so that $\pmb{\varepsilon}_i \perp \mathbf{e}_z$, $\mathbf{k}_i$. In the resonant elastic scattering process, the scattering cross-section $f_\mathrm{res}$ contains a term $\propto (\pmb{\varepsilon}_i \times \pmb{\varepsilon}_f) \cdot \mathbf{m}(\mathbf{q})$, where $\mathbf{q} = \mathbf{k}_f - \mathbf{k}_i$, and $\mathbf{m}(\mathbf{q})$ are the momentum transfer and the periodically modulated magnetic moment, respectively [Fig. \ref{Fig1}(c)]. In the present case, where $\pmb{\varepsilon}_i \propto \mathbf{k}_i \times \mathbf{e}_z$ and the scattering plane is aligned with the crystal's hexagonal $ab$ plane, we separate scattered x-rays with $\pmb{\varepsilon}_f \parallel \mathbf{e}_z \parallel \mathbf{c}^*$ (i.e., $\pi-\sigma^\prime$) and $\pmb{\varepsilon}_f \propto \mathbf{k}_f \times \mathbf{e}_z$ (i.e., $\pi-\pi^\prime$). Hence $f_\mathrm{res}^{\pi-\sigma^\prime}\propto \mathbf{k}_i \cdot  \mathbf{m}(\mathbf{q})$ and $f_\mathrm{res}^{\pi-\pi^\prime}\propto (\mathbf{k}_i \times \mathbf{k}_f) \cdot \mathbf{m}(\mathbf{q}) \propto m_z(\mathbf{q}) \sin(2\theta)$, where $2\theta$ is the scattering angle between $\mathbf{k}_i$ and $\mathbf{k}_f$. The observed scattered intensities are $I\propto \left| f_\mathrm{res} \right|^2$.

We obtain integrated intensities $I_{\pi-\sigma^\prime}$ and $I_{\pi-\pi^\prime}$ from Gaussian fits to line-cuts in momentum space. As described in the main text, the intensity ratio $R  = I_{\pi-\sigma^\prime} / I_{\pi-\pi^\prime}$ at a given magnetic reflection is sensitive to the cycloidal (proper screw) character of the magnetic order by virtue of being large (small) when $\mathbf{k}_i \cdot (\mathbf{ q} - \mathbf{G} )$ is large (small). For proper screw order, magnetic moments arrange themselves perpendicular to the propagation direction $(\mathbf{q}-\mathbf{G})$, with a finite projection onto both $\mathbf{e}_z$ (i.e., $m_{z}$) and the in-plane vector $\mathbf{e}_\perp = (\mathbf{q} - \mathbf{G}) \times \mathbf{e}_z$ (i.e., $\mathbf{m}_{ip}$). We introduced $\mathbf{G}$, the closest reciprocal lattice vector to $\mathbf{q}$, and $\mathbf{e}_z$, a unit vector along the $c$-direction. It follows $R \sin^2(2\theta) \propto \left[\mathbf{k}_i \cdot \mathbf{e}_\perp (\mathbf{q})\right]^2$ for a proper screw-type order, as demonstrated for Gd$_3$Ru$_4$Al$_{12}$ in Figs. \ref{Fig3}(b,c). Moreover, $R$ provides information on the degree of elliptical distortion, capturing e.g. the ratio $(m_y / m_z)^2$ for a proper screw propagating along $\mathbf{e}_x$, written as $\mathbf{e}_y \cdot m_y \sin(2\pi x / \lambda ) + \mathbf{e}_z \cdot  m_z \cos (2\pi x / \lambda)$, where $\mathbf{e}_x$, $\mathbf{e}_y$, $\mathbf{e}_z$ are Cartesian unit vectors aligned so that $\mathbf{e}_x \parallel \mathbf{q}$. In real materials with large (classical) magnetic moments, where the length of the moment is expected to be spatially uniform in space, this type of deformed screw can be realized by anharmonic distortion of the proper screw texture, as shown in Fig. \ref{Fig3}(b) (inset).

\section{Commensurate locking and Fourier modes}
\label{Esec:Fourier_modes}

\textbf{Spin Hamiltonian}\\
Considering a sum over a collection $\mathbf{r}_i$ of $N$ atomic sites (not necessarily periodic), we define the Fourier transform of the spin texture as~\cite{White07}
\begin{align}
\pmb{\mathcal{S}}(\mathbf{q}) &= \sum_{i} e^{i \mathbf{q}\cdot \mathbf{r}_i} \,\mathbf{S}_i\\
\mathbf{S}_j &= \frac{1}{N}\sum_{\mathbf{q}} e^{-i \mathbf{q}\cdot \mathbf{r}_j}\,\,\pmb{\mathcal{S}}(\mathbf{q}^\prime)
\end{align}
In the case of a periodic lattice, the atomic position can be decomposed into a unit cell coordinate, and an intra-cell coordinate viz. $\mathbf{r}_i = \mathbf{R}_i+\mathbf{d}_i$; the former satisfy $\exp\left(i \mathbf{G}\cdot \mathbf{R}_i\right)=1$ for all reciprocal lattice vectors $\mathbf{G}$; then, sums of exponentials over the lattice are factorized as
\begin{align}
\sum_i \exp\big(i \mathbf{r}_i \cdot \left(\mathbf{q}-\mathbf{q}^\prime\right)\big) &= \sum_{\mathbf{R}} \exp\big(i \mathbf{R} \cdot \left(\mathbf{q}-\mathbf{q}^\prime\right)\big) \sum_{\mathbf{d}\in\mathrm{u.c.}} \exp\big(i \mathbf{d} \cdot \left(\mathbf{q}-\mathbf{q}^\prime\right)\big)\notag\\
&= N_\mathrm{u.c.}\sum_{\mathbf{G}}\,\delta^{(3)}\left(\mathbf{q}-\mathbf{q}^\prime-\mathbf{G}\right)\,\sum_{\mathbf{d}\in\mathrm{u.c.}} \exp\big(i \mathbf{d} \cdot \mathbf{G}\big)
\label{Seq:factorize_delta}
\end{align}
and $N_\mathrm{u.c.}$ is the number of unit cells in the lattice. If $\mathbf{q}-\mathbf{q}^\prime = 0$, the latter factor is equal to the number of atoms in a unit cell.

We consider a spin Hamiltonian with Heisenberg exchange, magnetic anisotropy, and Zeeman energy,
\begin{equation}
    \mathcal{H} = \mathcal{H}_{ex}+\mathcal{H}_\mathrm{anis}+\mathcal{H}_Z = -2\sum_{i\neq j} J_{ij}\,\mathbf{S}_i\cdot\mathbf{S}_j - K_1 \sum_i \left(S_i^z\right)^2 + \mathbf{B}\cdot\sum_i \mathbf{S}_i
\end{equation}
Further assuming 
\begin{equation}
\label{Seq:Jij_distance}
J_{ij} = J(\mathbf{r}_i-\mathbf{r}_j)
\end{equation}
only depends on the distance between magnetic moments, a Fourier transform can be executed as  
\begin{align}
\mathcal{H}_{ex} &= -2\sum_{\mathbf{q}} J(\mathbf{q})\,\pmb{\mathcal{S}}(\mathbf{q})\cdot\pmb{\mathcal{S}}(-\mathbf{q})\\
J(\mathbf{q}) &= \frac{1}{N} \sum_{i\neq j} J(\mathbf{r}_i-\mathbf{r}_j)\, \exp(-\mathbf{q}\cdot(\mathbf{r}_i-\mathbf{r}_j))
\end{align}
where $\mathbf{q}$ is \textit{not} restricted to the first Brillouin zone. With additional exchange anisotropy in Fourier space, the expression is~\cite{Hirschberger21b}
\begin{equation}
\mathcal{H}_{ex} = -2\sum_{\mathbf{q}} J(\mathbf{q}) \sum_{\alpha\beta} \Gamma^{\alpha\beta}_\mathbf{q}\, \mathcal{S}^\alpha(\mathbf{q}) \mathcal{S}^\beta(-\mathbf{q})
\label{eq:hamiltonian2}
\end{equation}
and $\Gamma_\mathbf{q}$ is a dimensionless matrix capturing beyond-Heisenberg effects, such as dipolar interactions and some effects of spin-orbit interactions (Dzyaloshinskii-Moriya interactions, Kitaev exchange, etc.). This $\mathcal{H}_{ex}$ does not generate any pinning to the underlying lattice, as long as Eq. (\ref{Seq:Jij_distance}) remains valid.

For the single-ion part of the magnetic Hamiltonian, Umklapp terms can play a role. For example, the leading term relevant to the present work is
\begin{align}
\mathcal{H}_\mathrm{anis} &= -K_1 \sum_i \left(S_i^z\right)^2\notag\\
&= -\frac{K_1}{N} \sum_{\mathbf{G}}\sum_{\mathbf{q}\in\mathrm{BZ}} \mathcal{S}^z(\mathbf{q})\mathcal{S}^z(\mathbf{G}-\mathbf{q})\Big[\frac{1}{N_d}\sum_{\mathbf{d}\in\mathrm{u.c.}} e^{i\mathbf{G}\cdot\mathbf{d}} \Big]
\end{align}
according to Eq. (\ref{Seq:factorize_delta}). If a limited number of Fourier modes are selected by $J(\mathbf{q})$ in $\mathcal{H}_{ex}$, the coupling between these modes, between these modes and their higher harmonics, or between two higher harmonics of a primary mode, can contribute an energy that depends on the phase of the spin texture relative to the lattice. \\

\textbf{Application to Gd$_3$Ru$_4$Al$_{12}$}\\
In Eq. (\ref{eq:mag_skyrmion}), the skyrmion texture is generated by superposition of three proper screws. Subsequent normalization of the texture generates anharmonic terms, including
\begin{equation}
\mathbf{S}(\mathbf{r}) = \sum_{k=1}^3 \sum_{n=1}^\infty  \Big[\mathbf{e}_z a_n \cos\left(n\mathbf{q}_k\cdot (\mathbf{r}-\mathbf{r}_0)\right)+\mathbf{e}_{\perp k} b_n \sin\left(n\mathbf{q}_k\cdot(\mathbf{r} -\mathbf{r}_0) \right)\Big]
\label{eq:spin_texture}
\end{equation}
As compared to Eqs. (\ref{eq:mag_spiral}, \ref{eq:mag_skyrmion}), the phase $\phi_k$ is rewritten in terms of a reference coordinate $\mathbf{r}_0$; this assumes that the total sum of the three phases is fixed to $\sum_k \phi_k = 0$, for a skyrmion lattice. Multi-$\mathbf{q}$ textures also have Fourier modes of the type $\sim(\mathbf{q}_k + \mathbf{q}_l)$ ($l\neq k$); these do not make dominant contributions to the commensurate pinning, but the numerical calculation in Supplementary Fig. \ref{Fig_S8} includes all such subdominant terms as well.

Given site symmetry $mm2$ for the rare earth in the $P6_3/mmc$ structure of Gd$_3$Ru$_4$Al$_{12}$, we introduce two types of single-ion anisotropy,
\begin{align}
-K_1&\sum_i \left( S^z_i\right)^2
\label{eq:anisotropy_z}\\
-N_1^\mathrm{local}&\sum_i \left( \mathbf{S}^{ip}_i\cdot \mathbf{e}_i\right)^2
\label{eq:anisotropy_ip}
\end{align}
where again $\mathbf{r}_i = \mathbf{R}_i + \mathbf{d}_i$ and $\mathbf{e}_i \equiv \mathbf{e}_\mathbf{d}$ are local preference axes in the $ab$ plane, Supplementary Fig. \ref{FIG_S_magnetization_anisotropy}(c); $\mathbf{S}_{ip}$ is the in-plane part of the local spin.

Starting with Eq. (\ref{eq:spin_texture}) as a spin-Ansatz, and inserting into Eqs. (\ref{eq:anisotropy_z}, \ref{eq:anisotropy_ip}), we execute the sum over $\mathbf{R}_i$. Out of many product terms, the only survivors are those where $n\mathbf{q}_k$, $m \mathbf{q}_l$ sum to a reciprocal lattice vector $\mathbf{G}$. If this $\mathbf{G}=0$, there is no dependence of the final expression on the $\mathbf{d}$. Such terms are the same for incommensurate and commensurate states, and do not set a preferred alignment of the skyrmion core with respect to the underlying lattice, in the C case.

In Gd$_3$Ru$_4$Al$_{12}$, the C-SkL has $\mathbf{q}_1 = 0.25\,a^*\,\mathbf{e}_{a^*}$ and two equivalent vectors $\mathbf{q}_2$, $\mathbf{q}_3$ are generated by $120^\circ$ rotations. Numerically, the normalization of Eq. (\ref{eq:spin_texture}) creates higher harmonic intensities $a_n, b_n$ ($n>1$) on the order of $5-10\,\%$ of $a_1, b_1$. Focusing on product terms $\sim a_1 a_m$ and $\sim b_1 b_m$ ($m>1$), which give the dominant $\mathbf{d}$-dependent contributions, we have $\mathbf{q}_1 \pm m \mathbf{q}_1 = \mathbf{G}_1 = a^*\,\mathbf{e}_{a^*}$ (and symmetry equivalents). The relevant expressions are 
\begin{align}
-K_1\, &\frac{a_1 a_3}{2} \sum_{k=1}^3 \sum_{\mathbf{d}}  \cos\Big(\mathbf{G}_k\cdot\left(\mathbf{d}-\mathbf{r_0}\right)\Big)\\
N_1^\mathrm{local}\, &\frac{b_1 b_3}{2} \sum_{k = 1}^3 \sum_{\mathbf{d}} \gamma_{d,k}^2  \cos\Big(\mathbf{G}_k\cdot\left(\mathbf{d}-\mathbf{r_0}\right)\Big)
\end{align}
and $\gamma_{d,k} = (\mathbf{e}_\mathbf{d}\cdot \mathbf{e}_{\perp k})$. The similar functional form for $\mathbf{d}$-dependent terms $\sim K_1$ and $\sim N_1^\mathrm{local}$ suggests that -- if $K_1 \gg N_1^\mathrm{local}$, expected for this structure type and consistent with observation of both all-in-all-out and vortex-type spin structures tuned by small magnetic fields in Dy$_3$Ru$_4$Al$_{12}$~\cite{Gao19} -- a discussion based on $K_1$ alone can capture the key aspect of C-locking, to lowest order, in Gd$_3$Ru$_4$Al$_{12}$. 

\newpage
\begin{figure}[h]
\centering
\includegraphics[trim=0.0cm 7.2cm 0.cm 2.9cm,clip,width=1.0\textwidth]{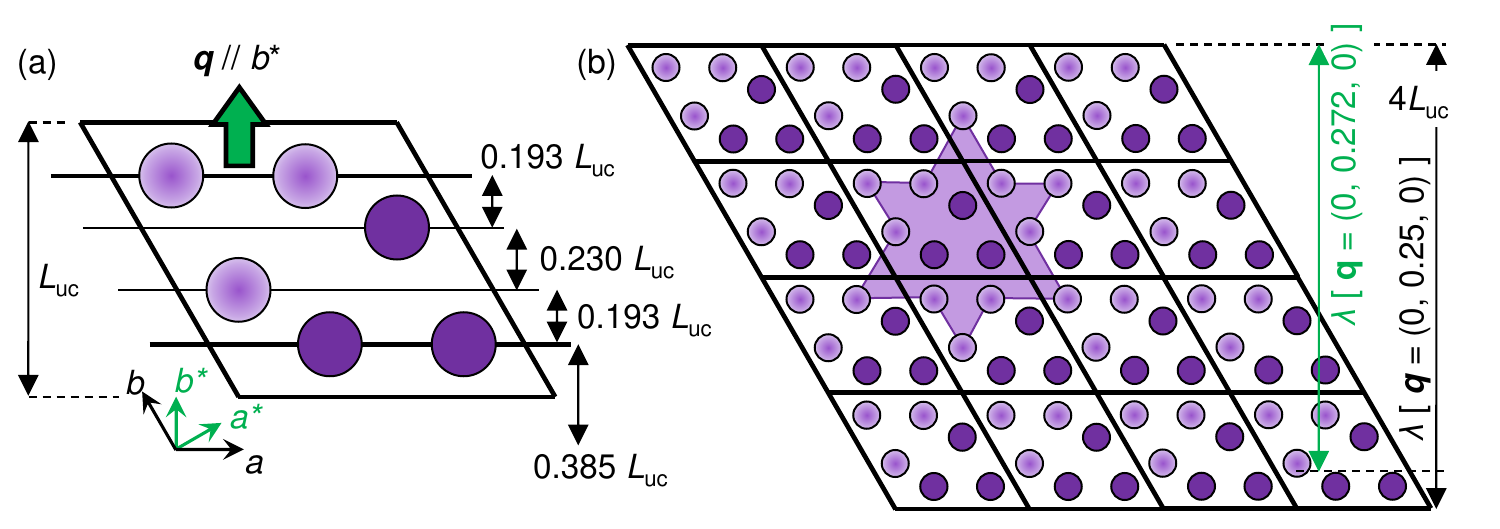}
\caption{Alignment of proper screw and skyrmion magnetic textures with respect to the crystal structure (magnetic sublattice) of Gd$_3$Ru$_4$Al$_{12}$ in hexagonal space group $P6_3/mmc$. (a) Crystallographic unit cell (UC) with upper (lower) trimer of Gd$^{3+}$ shown as purple (pink) circles. Al and Ru atoms are not shown, but crystal directions $a$, $b$, and reciprocal lattice axes $a^*$, $b^*$ are displayed at the bottom left side. The vertical arrow indicates $L_\mathrm{uc} = a\sin(60^\circ)$, the dimension of the UC projected parallel to the magnetic ordering vector $\mathbf{q}$ (green arrow). The nearest-neighbour distances between various rare earth planes perpendicular to $\mathbf{q}$ are also shown. (b) Magnetic sublattice within the magnetic UC ($4\times4$ crystallographic UCs). The kagome motif for the lower trimers is indicated by pink shading. The magnetic period $\lambda = 4\, L_{uc}$ corresponding to a hypothetical commensurate screw with $\mathbf{q} = (0.25, 0, 0)$ is indicated by black vertical arrows. Likewise, the green arrow shows a fractional $\lambda$ corresponding to the observed incommensurate proper screw (IC-PS) with $\mathbf{q} = (0.272, 0, 0)$.}\label{Fig_S9}
\end{figure}

\begin{figure}[h]
\centering
\includegraphics[width=1.0\textwidth]{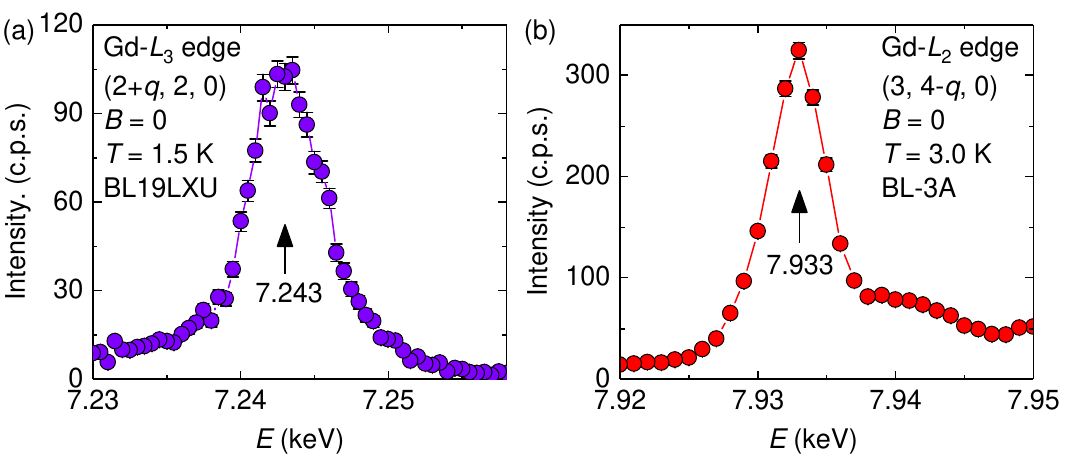}
\caption{Resonant enhancement of x-ray scattering intensity at the Gd-$L_{2,3}$ edges in the IC-PS (incommensurate proper screw) phase of Gd$_3$Ru$_4$Al$_{12}$. Data shown here is collected in reflection geometry, without analyser plate, while the sample is mounted in a cryomagnet and cooled to low temperature. Background from sample fluorescence has not been subtracted. The scattering plane was horizontal, and the incoming beam is $\pi$-polarized. (a) Gd-$L_3$ edge scattering at beamline BL19LXU of SPring-8 and (b) Gd-$L_2$ edge scattering at beamline BL-3A of Photon Factory (KEK) both exhibit sharp resonant enhancement. $(2+q,2, 0)$ and $(3, 4-q,0)$ with $q=0.272$ denote the momentum transfer $\mathbf{q}$ of the x-ray beam in reciprocal lattice units (r.l.u.). The data of this figure is recorded at a fixed position in momentum space, with $\pi$-polarized x-ray beam, but without analyser plate before the detector. }\label{Fig_S1}
\end{figure}

\begin{figure}[h]
\centering
\includegraphics[width=0.9\textwidth]{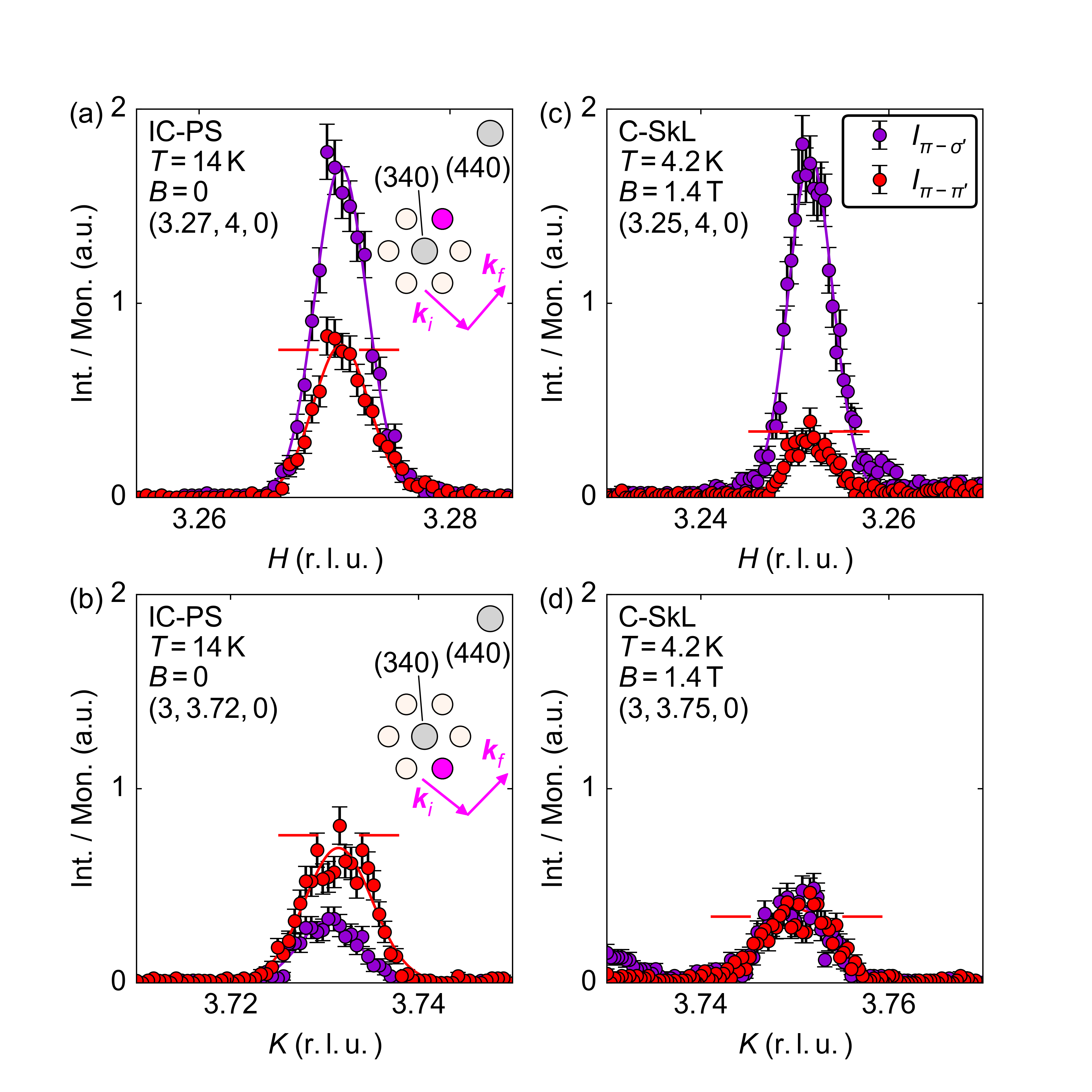}
\caption{Representative raw data of polarization analysis in REXS at the Gd-$L_{2}$ edge using beamline BL-3A of Photon Factory (KEK). IC-PS and C-SkL indicate incommensurate proper screw and commensurate skyrmion lattice phases, respectively. The purple and red symbols (lines) are scattering intensities (Gaussian fits) of data in the $\pi-\sigma^\prime$ and $\pi-\pi^\prime$ channels. We show red horizontal lines at $y = 0.76$ (at $y = 0.34$) in panels (a,b) [(c,d)] to indicate  comparable intensities of $I_{\pi-\pi^\prime}$ at different positions in momentum space (Note: A more quantitative comparison of absolute intensities at \textit{different} positions in momentum-space requires calculation of integrated intensities along $\theta$-$2\theta$ scans, consideration of the structure factor, corrections for the penetration depth of x-rays in magnetic scattering, etc.). The inset cartoons illustrate the scattering geometry in reciprocal space, where $\mathbf{k}_i$ and $\mathbf{k}_f$ are the incoming and outgoing beam vectors while grey, yellow, and pink circles indicate the fundamental Bragg reflection, six-fold symmetric magnetic satellites, and the magnetic satellite under consideration, respectively. Error bars are Poisson errors of the number of detector counts collected for each configuration of the diffractometer (Supplementary Section \ref{Esec:rexs_experimental}).}\label{Fig_SI_polarization_analysis}
\end{figure}

\begin{figure}[h]
\centering
\includegraphics[trim=0.9cm 0.0cm 0.cm 0.cm,clip,width=1.0\textwidth]{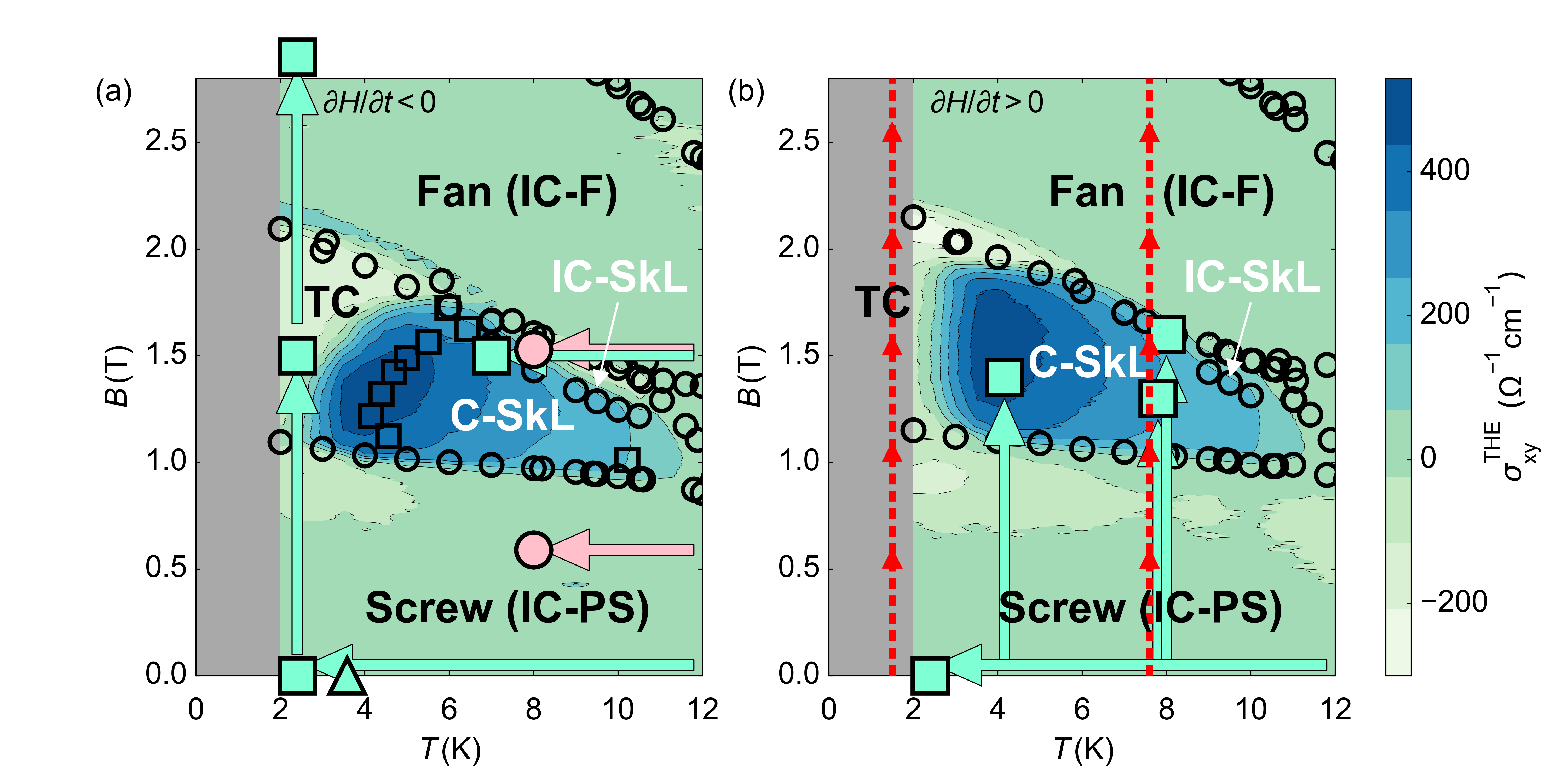}
\caption{Sample preparation history for resonant elastic x-ray scattering (REXS), Lorentz transmission electron microscopy (L-TEM), and neutron scattering measurements, comparing prior work and the present observation of a commensurate skyrmion lattice. Phase boundaries are determined from magnetization measurements, and the color map depicts the unconventional contribution to the Hall conductivity $\sigma_{xy}$~\cite{Hirschberger19}. The combination of L-TEM~\cite{Hirschberger19} and precise measurements of the Hall effect~\cite{Hirschberger21b} supports our identification of C-SkL and IC-SkL as multi-$\mathbf{q}$ (superposition) states. (a) The history of measurements in Ref. \cite{Hirschberger19}, and (b) the history of measurements in the present work, using a superior single crystal where locking between the skyrmion lattice and the crystal lattice could be clearly detected by x-rays. Green squares, green triangles, pink circles, and red lines with arrows correspond to REXS measurements with polarization analysis, elastic neutron scattering, L-TEM measurements, and REXS measurements for line-scans (Figs. \ref{Fig1}, \ref{Fig3} of the main text), respectively. For symbols, the sample history (field cooling FC, zero-field cooling ZFC, and so on) is indicated by fat arrows. The contour map, corresponding to the magnitude of the topological Hall conductivity $\sigma_{xy}^\mathrm{THE}$, as well as phase boundaries from magnetization and transport experiments (black open circles) are adapted from Ref. \cite{Hirschberger19}. The previously discussed incommensurate skyrmion lattice is denoted as IC-SkL. Note that phase boundaries and contour maps in left (right) panels are obtained from decreasing (increasing) field scans, respectively. As elsewhere in this manuscript, $B = \mu_0 (H_\mathrm{ext} - NM)$ with magnetization $M$ denotes the magnetic field after demagnetization correction.}\label{Fig_SI_phase_diagram}
\end{figure}

\begin{figure}[h]
\centering
\includegraphics[trim=0.0cm 1.8cm 0.cm 0.cm,clip,width=1.0\textwidth]{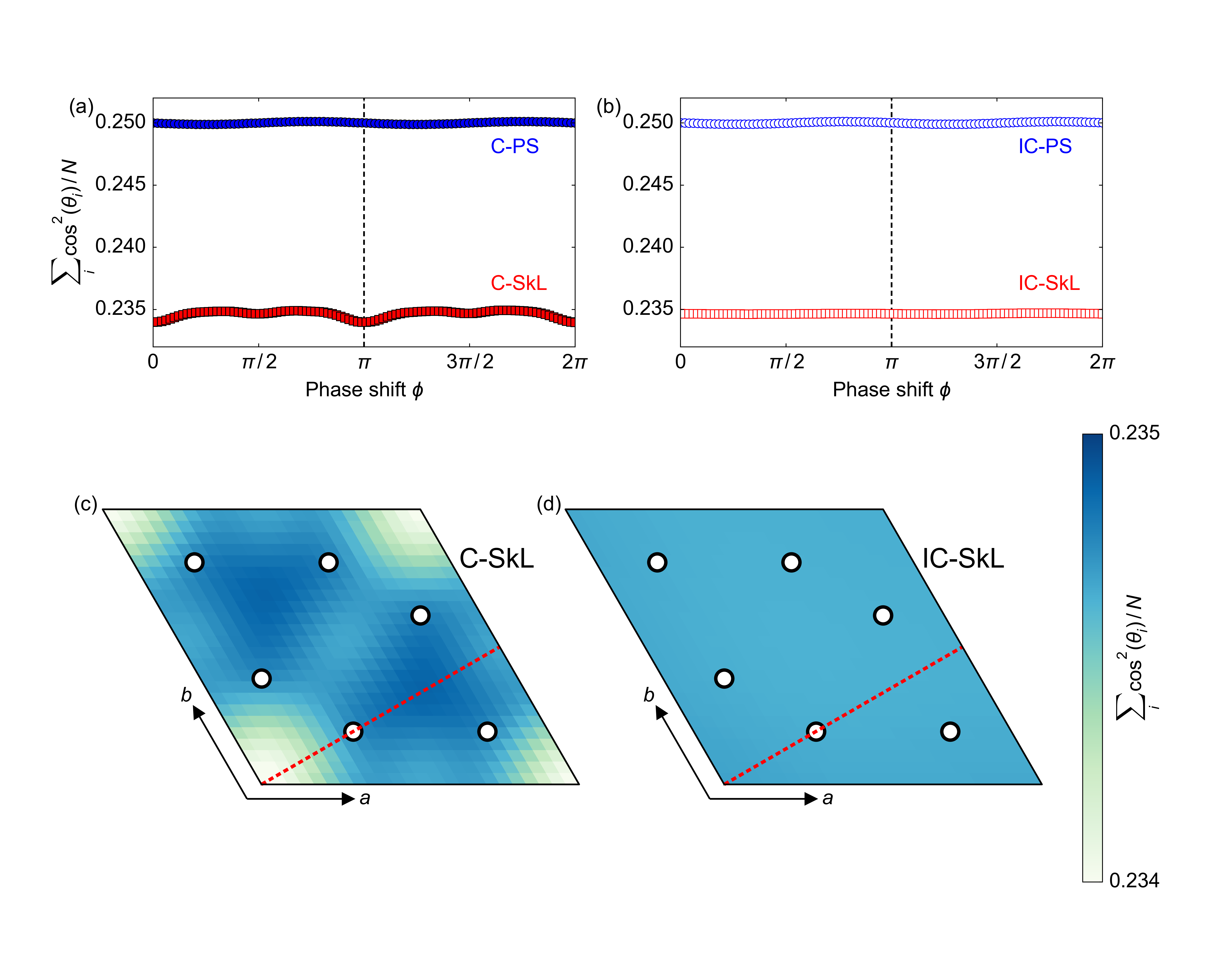}
\caption{Estimation of anisotropy energy for commensurate proper screw, commensurate skyrmion lattice (C-PS, C-SkL) and incommensurate proper screw, incommensurate skyrmion lattice (IC-PS, IC-SkL) on the breathing kagome lattice of Gd$_3$Ru$_4$Al$_{12}$. Consider the anisotropy term in the Hamiltonian, $A\sum_i \left(S_i^z\right)^2$, a sum over lattice sites $i$. Easy-plane anisotropy $A>0$ favors small $\sum_i \cos^2(\theta_i)/N$, shown here for various magnetic textures; the angles $\theta_i$ denote the $z$-projection of magnetic moments on $N = 1499\times 1499 \times 6$ lattice sites, given $6$ magnetic sites in a UC. (a) Blue: the total $z$-projection of moments, normalized to $N$, for C-PS as in Eq. (\ref{eq:mag_spiral}) with $\mathbf{q}_1=0.25\cdot \mathbf{e}_{a^*}$ and $\phi \in [0, 2 \pi]$. Red: The same for C-SkL as in Eq. (\ref{eq:mag_skyrmion}) with three ordering vectors obeying $\left|\mathbf{q}_i \right| = 0.25\, \mathbf{e}_{a^*}$ and phases $\phi_1=+\phi$, $\phi_2=\phi_3=-\phi/2$. (b) The same for IC-PS, IC-SkL with $\left|\mathbf{q}_i \right| = 0.272\, a^*$. (c,d) Sum of squared $z$-projection of moments, normalized to $N$, for C-SkL and IC-SkL with $\left|\mathbf{q}_i \right| = 0.25\, a^*$, $0.272\, a^*$, respectively. The color of a pixel at $(x,y)$ corresponds to the sum of $\cos^2(\theta_i)$ over the lattice, if the skyrmion core is located at $(x,y)$. The black outer frame and black circles with white filling signify a crystallographic UC and the positions of magnetic Gd$^{3+}$ sites, projected onto a single $z=0$ plane, respectively. Red dashed lines denote the zero-position of the magnetic proper screw (core position of the skyrmion) considered in panels (a,b). Note that $\phi=0$ implies the skyrmion core is at the $(0,0)$ position at the lower left side of the contour maps.}\label{Fig_S8}
\end{figure}

\begin{figure}[htb]
\centering
\includegraphics[trim=0.0cm 1.6cm 0.cm 1.6cm, clip,width=0.9\textwidth]{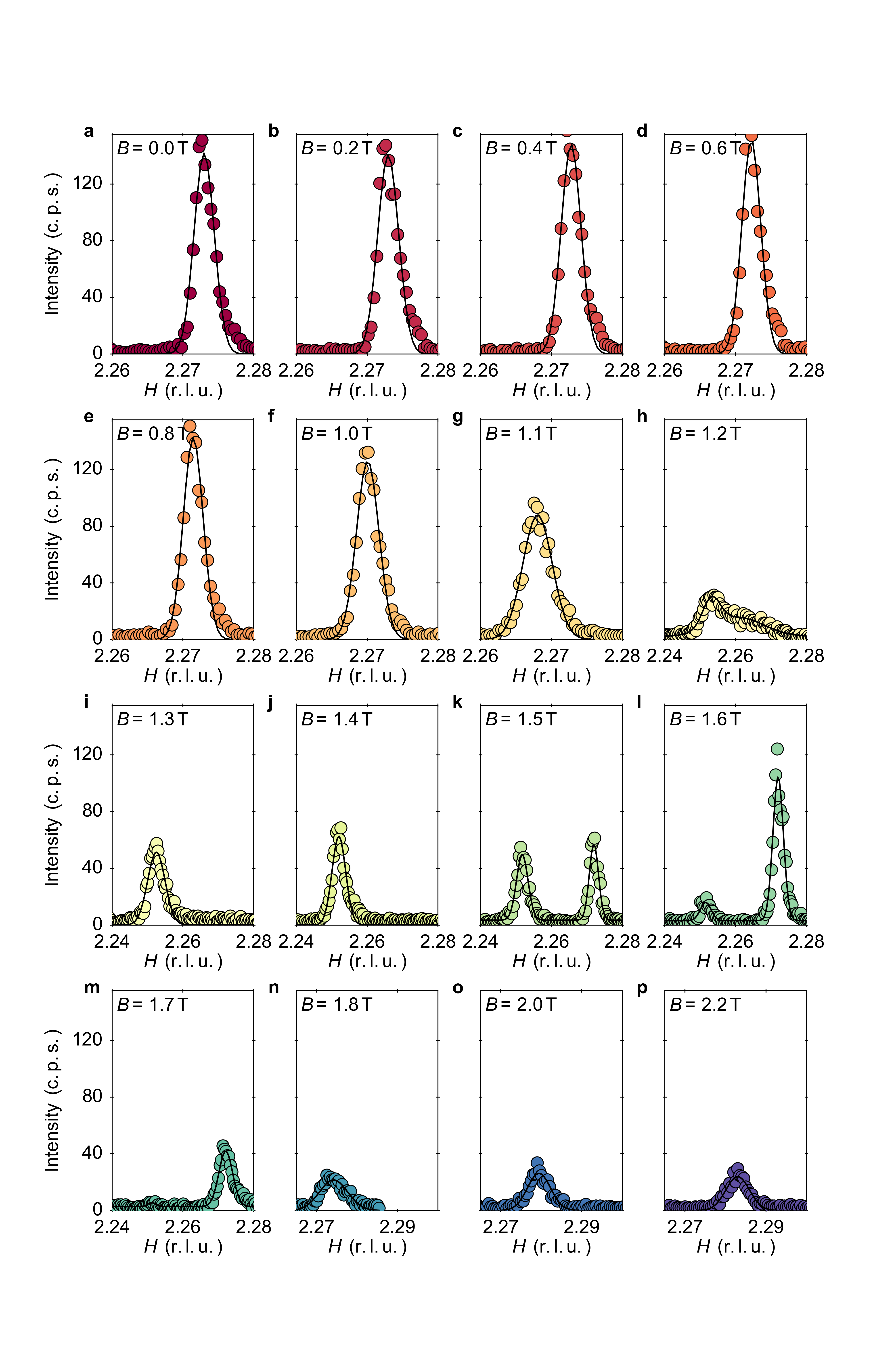}
\caption{Line-scan profiles of resonant x-ray scattering at $(H, 2, 0)$ and $T = 7.6\,$K, constituting the raw data for the colormap in Fig. \ref{Fig1}. The scattering intensity in counts per second (c.p.s.) is shown as a function of the Miller index $H$, for various magnetic field values ($\mathbf{B}\parallel c$) and otherwise unchanged sample condition. The data were collected in zero-field cooled condition. Gaussian fit profiles are shown as black lines. For some magnetic field, a double-peak profile indicates phase coexistence of two magnetic scattering vectors. Data obtained without analyser crystal at BL19LXU of SPring-8, without analyser plate. }\label{Fig_S_Bdep_7p6K}
\end{figure}

\begin{figure}[htb]
\centering
\includegraphics[trim=0.0cm 0.6cm 0.cm 0.cm,clip,width=0.97\textwidth]{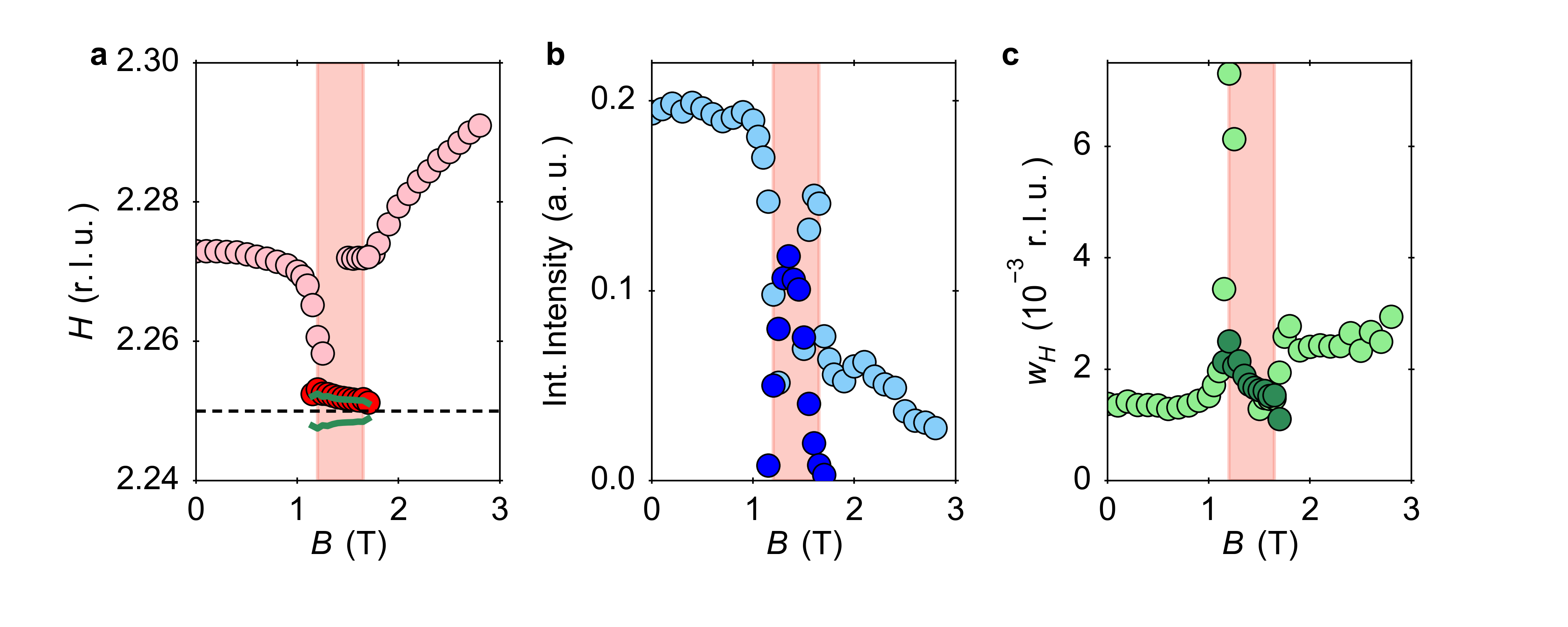}
\caption{Gaussian fit parameters for line-scan profiles in resonant elastic x-ray scattering at $T = 7.6\,$K and $(H, 2, 0)$, corresponding to black profiles in Supplementary Fig. \ref{Fig_S_Bdep_7p6K}. \textbf{a}, Peak center of Gaussian profile, with red and light red symbols indicating commensurate (C) and incommensurate (IC) reflections, respectively. The value $H_0 = 2.25$ is indicated by a horizontal black dashed line, and $H_0 \pm w_H$ with $w_H$ from panel (c) is denoted by dark green lines. As $w_H$ of the C reflection decreases with rising $B$, notably, the center of the reflection approaches more closely to $H_0$. \textbf{b}, Integrated peak intensity (from $H$-scans) in arbitrary units (a.u.); blue and light blue symbols are C and IC intensities. \textbf{c}, Gaussian peak width $w_H$, in units of reciprocal lattice units (r.l.u.). Deep ad light green symbols show width of C and IC reflections, respectively. In all panels, red highlight marks the C-SkL phase. This experiment was carried out at BL19LXU of SPring-8, without analyser plate. }\label{Fig_S_Bscan_fitparams_7p6K}
\end{figure}

\begin{figure}[htb]
\centering
\includegraphics[trim=0.0cm 0.6cm 0.cm 0.cm,clip,width=0.97\textwidth]{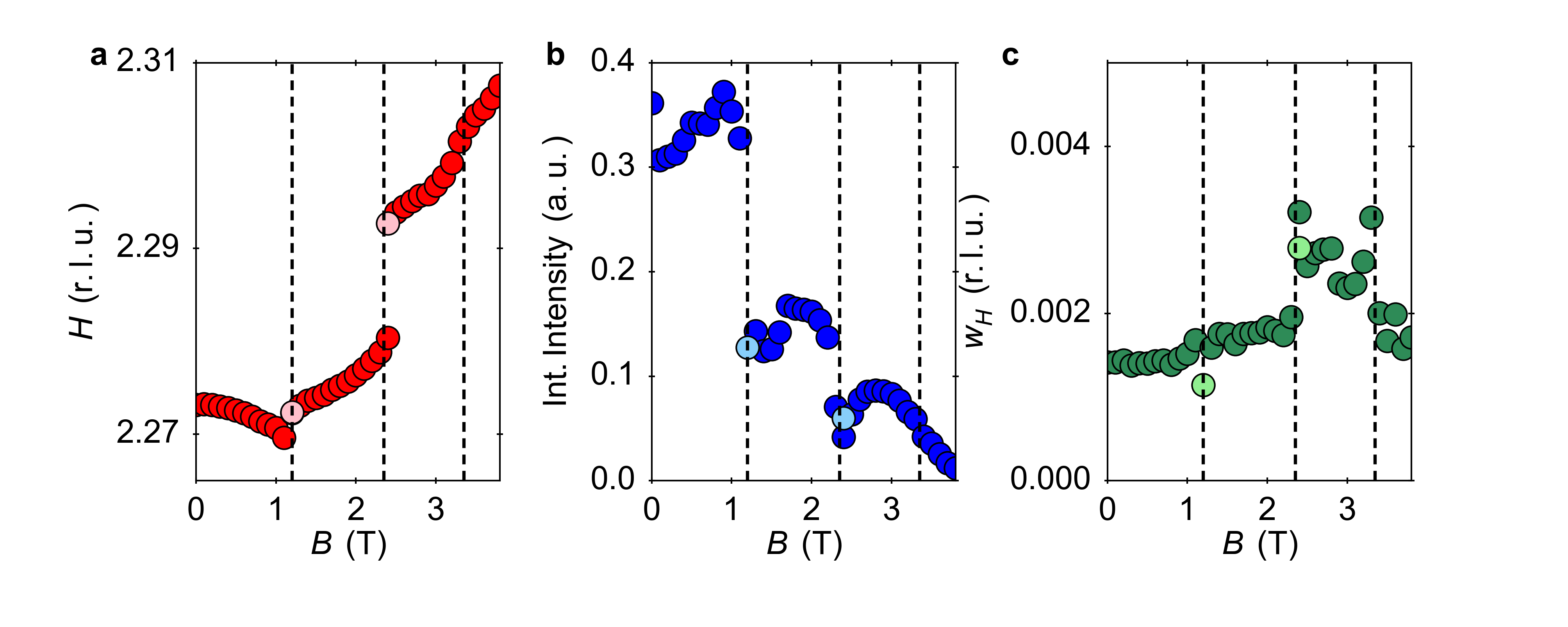}
\caption{Gaussian fit parameters for line-scan profiles in resonant elastic x-ray scattering at $T = 1.5\,$K and $(H, 2, 0)$. \textbf{a}, Peak center of Gaussian profile. \textbf{b}, Integrated peak intensity (from $H$-scans) in arbitrary units (a.u.). \textbf{c}, Gaussian peak width $w_H$, in units of reciprocal lattice units (r.l.u.). In the regime where a double-peak profile was fitted, corresponding to phase coexistence at first-order phase boundaries, symbols are shown in two different colors -- e.g. dark and light red in panel (a). Dashed vertical lines separate magnetic phases: From left to right, IC-PS, IC-TC, IC-Fan, and an unidentified (possibly fan) phase that was first noted in Ref. \cite{Hirschberger19}. This experiment was carried out at BL19LXU of SPring-8, without analyser plate. }\label{Fig_S_Bscan_fitparams_1p5K}
\end{figure}

\begin{figure}[htb]
\centering
\includegraphics[trim=0.0cm 0.6cm 0.cm 0.cm,clip,width=0.97\textwidth]{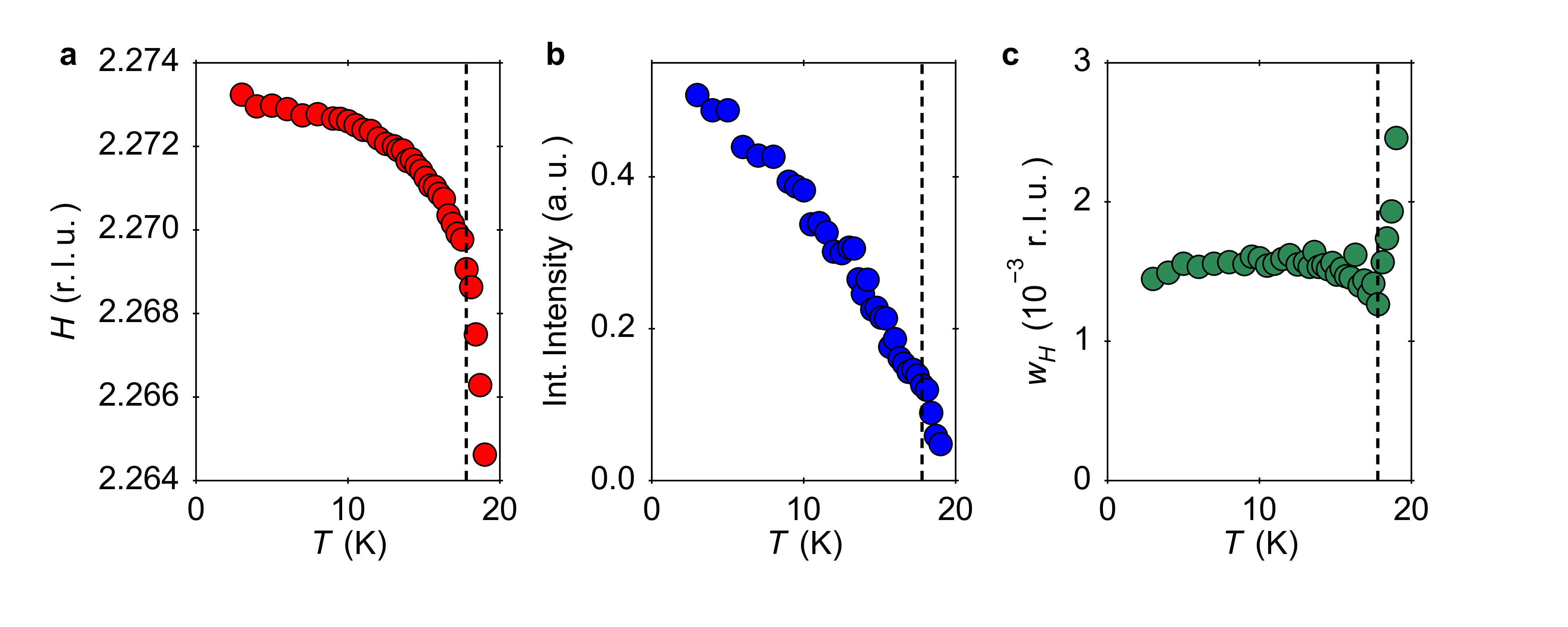}
\caption{Gaussian fit parameters for line-scan profiles in resonant elastic x-ray scattering at $B = 0\,$T, as a function of temperature, for the reflection at $(H, 2, 0)$. \textbf{a}, Peak center of Gaussian profile. \textbf{b}, Integrated peak intensity (from $H$-scan) in arbitrary units (a.u.). \textbf{c}, Gaussian peak width $w_H$, in units of reciprocal lattice units (r.l.u.). Dashed vertical lines separate two magnetic phases: From left to right, IC-PS and IC-sinusoid / fan-type. This experiment was carried out at BL19LXU of SPring-8, without analyser plate. }\label{Fig_S_Tscan_fitparams_0T}
\end{figure}

\begin{figure}[htb]
\centering
\includegraphics[trim=0.0cm 0.6cm 0.cm 0.cm,clip,width=0.97\textwidth]{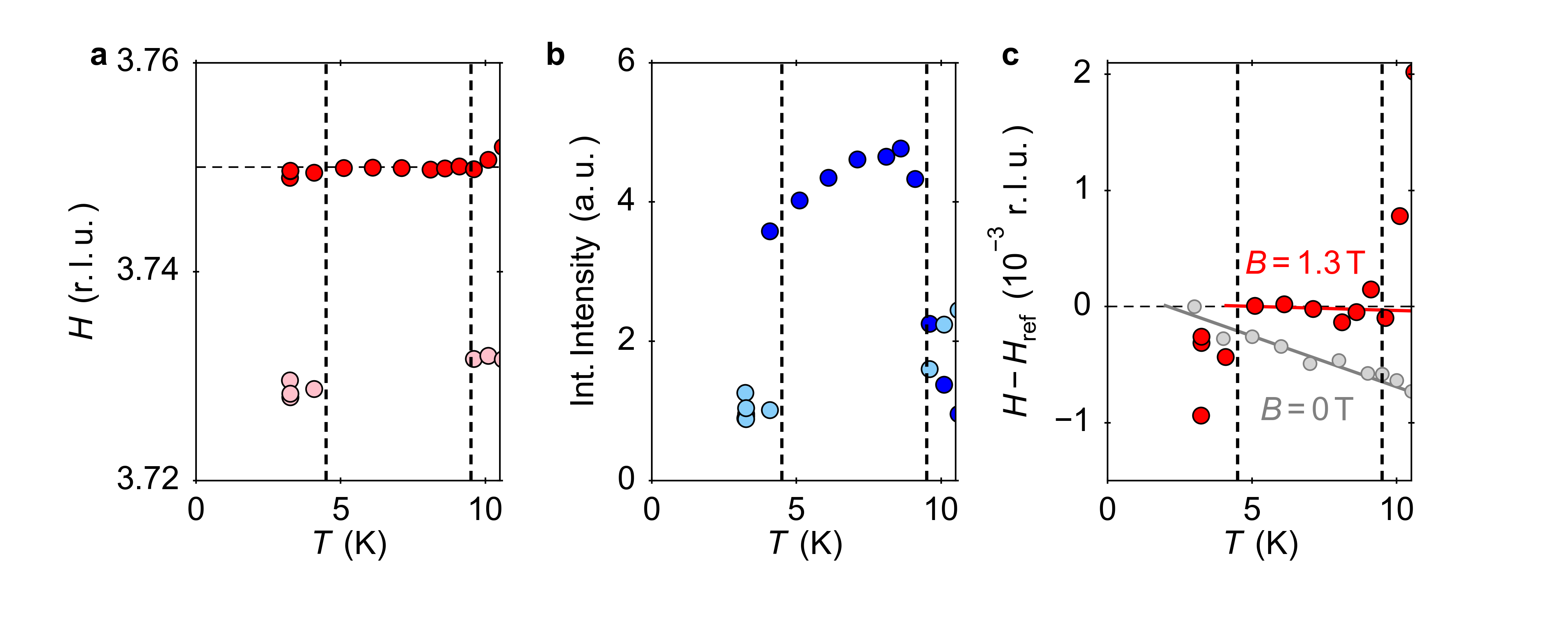}
\caption{Gaussian fit parameters for line-scan profiles in resonant elastic x-ray scattering at $B = 1.3\,$T, as a function of temperature, from $(4-\delta, 3+\delta, 0)$-scans. \textit{Note}: these one-dimensional scans are along the line connecting the $(4,3,0)$ and $(3,4,0)$ Bragg reflections in reciprocal space. \textbf{a}, Peak center of Gaussian profile, where $H = 4-\delta$. Dark and light colored symbols indicate intensity from commensurate and incommensurate magnetic reflections, respectively. In the present sample, phase coexistence of C-SkL and IC-TC was observed at base temperature, despite field cooling. Dashed vertical lines separate magnetic phases: From left to right, IC-TS, C-SkL, IC-Fan. \textbf{b}, Integrated peak intensity for commensurate (dark blue) and incommensurate (light blue) reflections in arbitrary units (a.u.), from abovementioned $HK$-scans. At the lowest temperature, the intensities  \textbf{c}, Weak temperature dependence of ordering vector in C-SkL, when compared to IC-PS. We show center positions of Gaussian profiles from panel (a) and Supplementary Fig. \ref{Fig_S_Tscan_fitparams_0T}(a) in red and grey, respectively. The data are shifted by $H_\mathrm{ref}$, which corresponds to the $H$-value at $T = 3\,$K and $T = 5\,$K for the zero-field and in-field data, respectively. Lines are linear fits to the data at $5\,\mathrm{K}<T<10\,\mathrm{K}$ and over the entire visible temperature range, respectively. This experiment was carried out at BL-3A of Photon Factory, in $\pi$-$\sigma^\prime$ geometry with PG-006 analyser. }\label{Fig_S_Tscan_fitparams_1p3T}
\end{figure}

\begin{figure}[htb]
\centering
\includegraphics[trim=0.0cm 0.0cm 0.cm 0.cm,clip,width=0.75\textwidth]{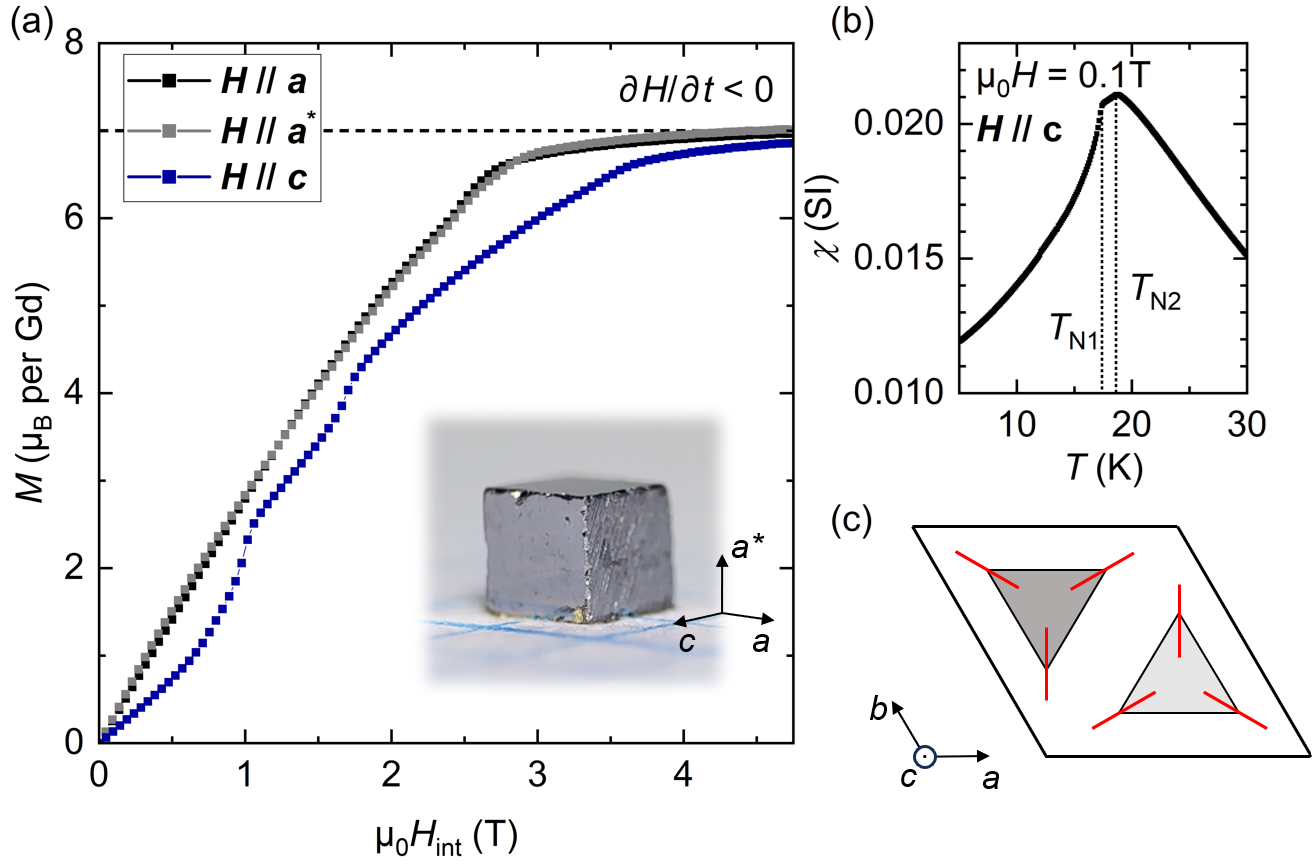}
\caption{Magnetization anisotropy of a single crystal of Gd$_3$Ru$_4$Al$_{12}$ at $T = 2\,$K. \textbf{a}, Magnetization isotherms $
M(H)$ for three crystalline directions. Dashed horizontal line indicates expected saturation magnetisation, $7\,\mu_B/\mathrm{Gd}$. The curves for two in-plane directions coincide nearly perfectly; c.f. Fig. 6 of Ref. \cite{Nakamura23}. Consistent with the FMR study in the main text, the saturation field is about $1\,$T lower for magnetic field in the basal plane. Inset: single crystalline sample, hand-polished into a cube of dimensions $1\times 1\times 1\,$mm$^3$. The three principal directions were aligned by Laue x-ray scattering before polishing. \textbf{b}, Temperature dependent magnetization at external field $\mu_0 H_\mathrm{ext} = 0.1\,$T. Two critical temperatures, also reported in Ref. \cite{Hirschberger19}, are indicated by dashed vertical lines. \textbf{c}, Sketch of the hexagonal unit cell of Gd$_3$Ru$_4$Al$_{12}$, projected onto the $z = 0$ plane. Only gadolinium trimers are shown (shaded triangles) with local mirror planes at the rare earth site. The mirror planes are spanned by the red lines and the $c$-axis, where the unit vectors $\mathbf{e}_\mathbf{d}$ in section \ref{Esec:Fourier_modes} are aligned with the red line for each site $\mathbf{d}$, respectively.}\label{FIG_S_magnetization_anisotropy}
\end{figure}

\begin{table}[h!]
    \centering 
    \caption{\textbf{Raw data of polarization analysis in magnetic x-ray scattering, in phase C-SkL}. This data was used for the determination of spin directions in Fig. \ref{Fig3}.}
    \label{Etable:CSkL_intensity}
    \vspace*{1em}
    \begin{tabular}{|c|c|c|c|c|c|c|}
    \hline
    \hline
    \hspace*{.7em} $H$  \hspace*{.7em} & \hspace*{.7em} $K$ \hspace*{.7em} & \hspace*{.7em} $L$ \hspace*{.7em} & \hspace*{1.4em} $2\theta$ \hspace*{1.4em} & \hspace*{.7em} $\left[\mathbf{k}_i\cdot\mathbf{e}_\perp(\mathbf{q})\right]^2$\hspace*{.7em} & \hspace*{.7em} $R\sin^2(2\theta)$\hspace*{.7em} & \hspace*{.7em} $R$\hspace*{.7em}\\
    \hline
    \multicolumn{3}{|c|}{r.l.u.} & degrees & -- &  -- & --\\
    \hline
    \hline

    $2.75$  &  $4.25$  &  $0.0$  &  $77$  &  $0.27$  &  $3.33$  &  $3.5$  \\
    $8.0$  &  $-0.75$  &  $0.0$  &  $102$  &  $0.94$  &  $6.79$  &  $7.12$  \\
    $7.75$  &  $-0.75$  &  $0.0$  &  $97$  &  $0.06$  &  $0.74$  &  $0.75$  \\
    $7.75$  &  $-1.0$  &  $0.0$  &  $96$  &  $0.56$  &  $5.24$  &  $5.3$  \\
    $4.75$  &  $2.25$  &  $0.0$  &  $78$  &  $0.21$  &  $2.05$  &  $2.13$  \\
    $4.75$  &  $2.0$  &  $0.0$  &  $75$  &  $0.34$  &  $1.96$  &  $2.09$  \\
    $5.0$  &  $2.25$  &  $0.0$  &  $81$  &  $1.0$  &  $6.02$  &  $6.14$  \\
    $2.75$  &  $4.25$  &  $0.0$  &  $77$  &  $0.27$  &  $4.79$  &  $5.05$  \\
    $2.75$  &  $4.0$  &  $0.0$  &  $73$  &  $0.24$  &  $0.53$  &  $0.58$  \\
    $3.0$  &  $4.25$  &  $0.0$  &  $80$  &  $0.99$  &  $7.58$  &  $7.82$  \\
    $2.0$  &  $2.25$  &  $0.0$  &  $44$  &  $0.97$  &  $5.84$  &  $12.07$  \\
    $2.25$  &  $2.0$  &  $0.0$  &  $44$  &  $0.41$  &  $1.85$  &  $3.81$  \\
    $7.0$  &  $0.75$  &  $0.0$  &  $97$  &  $0.99$  &  $9.15$  &  $9.33$  \\
    $7.25$  &  $1.0$  &  $0.0$  &  $105$  &  $0.26$  &  $0.23$  &  $0.24$  \\

    \hline
    \hline
    \end{tabular}
\end{table}

\begin{thebibliography}{48}%
\makeatletter
\providecommand \@ifxundefined [1]{%
 \@ifx{#1\undefined}
}%
\providecommand \@ifnum [1]{%
 \ifnum #1\expandafter \@firstoftwo
 \else \expandafter \@secondoftwo
 \fi
}%
\providecommand \@ifx [1]{%
 \ifx #1\expandafter \@firstoftwo
 \else \expandafter \@secondoftwo
 \fi
}%
\providecommand \natexlab [1]{#1}%
\providecommand \enquote  [1]{``#1''}%
\providecommand \bibnamefont  [1]{#1}%
\providecommand \bibfnamefont [1]{#1}%
\providecommand \citenamefont [1]{#1}%
\providecommand \href@noop [0]{\@secondoftwo}%
\providecommand \href [0]{\begingroup \@sanitize@url \@href}%
\providecommand \@href[1]{\@@startlink{#1}\@@href}%
\providecommand \@@href[1]{\endgroup#1\@@endlink}%
\providecommand \@sanitize@url [0]{\catcode `\\12\catcode `\$12\catcode
  `\&12\catcode `\#12\catcode `\^12\catcode `\_12\catcode `\%12\relax}%
\providecommand \@@startlink[1]{}%
\providecommand \@@endlink[0]{}%
\providecommand \url  [0]{\begingroup\@sanitize@url \@url }%
\providecommand \@url [1]{\endgroup\@href {#1}{\urlprefix }}%
\providecommand \urlprefix  [0]{URL }%
\providecommand \Eprint [0]{\href }%
\providecommand \doibase [0]{https://doi.org/}%
\providecommand \selectlanguage [0]{\@gobble}%
\providecommand \bibinfo  [0]{\@secondoftwo}%
\providecommand \bibfield  [0]{\@secondoftwo}%
\providecommand \translation [1]{[#1]}%
\providecommand \BibitemOpen [0]{}%
\providecommand \bibitemStop [0]{}%
\providecommand \bibitemNoStop [0]{.\EOS\space}%
\providecommand \EOS [0]{\spacefactor3000\relax}%
\providecommand \BibitemShut  [1]{\csname bibitem#1\endcsname}%
\let\auto@bib@innerbib\@empty
\bibitem [{\citenamefont {Bogdanov}\ and\ \citenamefont
  {Yablonsky}(1989)}]{Bogdanov89}%
  \BibitemOpen
  \bibfield  {author} {\bibinfo {author} {\bibfnamefont {A.}~\bibnamefont
  {Bogdanov}}\ and\ \bibinfo {author} {\bibfnamefont {D.}~\bibnamefont
  {Yablonsky}},\ }\bibfield  {title} {\bibinfo {title} {{Thermodynamically
  stable vortexes in magnetically ordered crystals: mixed state of
  magnetics}},\ }\href@noop {} {\bibfield  {journal} {\bibinfo  {journal}
  {Journal of Experimental and Theoretical Physics}\ }\textbf {\bibinfo
  {volume} {95}},\ \bibinfo {pages} {178} (\bibinfo {year} {1989})}\BibitemShut
  {NoStop}%
\bibitem [{\citenamefont {M{\"u}hlbauer}\ \emph {et~al.}(2009)\citenamefont
  {M{\"u}hlbauer}, \citenamefont {Binz}, \citenamefont {Jonietz}, \citenamefont
  {Pfleiderer}, \citenamefont {Rosch}, \citenamefont {Neubauer}, \citenamefont
  {Georgii},\ and\ \citenamefont {B{\"o}ni}}]{Muehlbauer09}%
  \BibitemOpen
  \bibfield  {author} {\bibinfo {author} {\bibfnamefont {S.}~\bibnamefont
  {M{\"u}hlbauer}}, \bibinfo {author} {\bibfnamefont {B.}~\bibnamefont {Binz}},
  \bibinfo {author} {\bibfnamefont {F.}~\bibnamefont {Jonietz}}, \bibinfo
  {author} {\bibfnamefont {C.}~\bibnamefont {Pfleiderer}}, \bibinfo {author}
  {\bibfnamefont {A.}~\bibnamefont {Rosch}}, \bibinfo {author} {\bibfnamefont
  {A.}~\bibnamefont {Neubauer}}, \bibinfo {author} {\bibfnamefont
  {R.}~\bibnamefont {Georgii}},\ and\ \bibinfo {author} {\bibfnamefont
  {P.}~\bibnamefont {B{\"o}ni}},\ }\bibfield  {title} {\bibinfo {title}
  {{Skyrmion Lattice in a Chiral Magnet}},\ }\href@noop {} {\bibfield
  {journal} {\bibinfo  {journal} {Science}\ }\textbf {\bibinfo {volume}
  {323}},\ \bibinfo {pages} {915} (\bibinfo {year} {2009})}\BibitemShut
  {NoStop}%
\bibitem [{\citenamefont {Tokura}\ and\ \citenamefont
  {Kanazawa}(2021)}]{Tokura21}%
  \BibitemOpen
  \bibfield  {author} {\bibinfo {author} {\bibfnamefont {Y.}~\bibnamefont
  {Tokura}}\ and\ \bibinfo {author} {\bibfnamefont {N.}~\bibnamefont
  {Kanazawa}},\ }\bibfield  {title} {\bibinfo {title} {{Magnetic Skyrmion
  Materials}},\ }\href@noop {} {\bibfield  {journal} {\bibinfo  {journal}
  {Chemical Reviews}\ }\textbf {\bibinfo {volume} {5}},\ \bibinfo {pages}
  {2857--2897} (\bibinfo {year} {2021})}\BibitemShut {NoStop}%
\bibitem [{\citenamefont {Binz}\ \emph {et~al.}(2006)\citenamefont {Binz},
  \citenamefont {Vishwanath},\ and\ \citenamefont {Aji}}]{Binz06}%
  \BibitemOpen
  \bibfield  {author} {\bibinfo {author} {\bibfnamefont {B.}~\bibnamefont
  {Binz}}, \bibinfo {author} {\bibfnamefont {A.}~\bibnamefont {Vishwanath}},\
  and\ \bibinfo {author} {\bibfnamefont {V.}~\bibnamefont {Aji}},\ }\bibfield
  {title} {\bibinfo {title} {{Theory of the Helical Spin Crystal: A Candidate
  for the Partially Ordered State of MnSi}},\ }\href@noop {} {\bibfield
  {journal} {\bibinfo  {journal} {Physical Review Letters}\ }\textbf {\bibinfo
  {volume} {96}},\ \bibinfo {pages} {207202} (\bibinfo {year}
  {2006})}\BibitemShut {NoStop}%
\bibitem [{\citenamefont {Binz}\ and\ \citenamefont
  {Vishwanath}(2006)}]{Binz06b}%
  \BibitemOpen
  \bibfield  {author} {\bibinfo {author} {\bibfnamefont {B.}~\bibnamefont
  {Binz}}\ and\ \bibinfo {author} {\bibfnamefont {A.}~\bibnamefont
  {Vishwanath}},\ }\bibfield  {title} {\bibinfo {title} {{Theory of helical
  spin crystals: Phases, textures, and properties}},\ }\href@noop {} {\bibfield
   {journal} {\bibinfo  {journal} {Physical Review B}\ }\textbf {\bibinfo
  {volume} {74}},\ \bibinfo {pages} {214408} (\bibinfo {year}
  {2006})}\BibitemShut {NoStop}%
\bibitem [{\citenamefont {Okubo}\ \emph {et~al.}(2012)\citenamefont {Okubo},
  \citenamefont {Chung},\ and\ \citenamefont {Kawamura}}]{Okubo12}%
  \BibitemOpen
  \bibfield  {author} {\bibinfo {author} {\bibfnamefont {T.}~\bibnamefont
  {Okubo}}, \bibinfo {author} {\bibfnamefont {S.}~\bibnamefont {Chung}},\ and\
  \bibinfo {author} {\bibfnamefont {H.}~\bibnamefont {Kawamura}},\ }\bibfield
  {title} {\bibinfo {title} {{Multiple-q States and the Skyrmion Lattice of the
  Triangular-Lattice Heisenberg Antiferromagnet under Magnetic Fields}},\
  }\href@noop {} {\bibfield  {journal} {\bibinfo  {journal} {Physical Review
  Letters}\ }\textbf {\bibinfo {volume} {108}},\ \bibinfo {pages} {017206}
  (\bibinfo {year} {2012})}\BibitemShut {NoStop}%
\bibitem [{\citenamefont {Hayami}\ \emph {et~al.}(2016)\citenamefont {Hayami},
  \citenamefont {Lin},\ and\ \citenamefont {Batista}}]{Hayami16}%
  \BibitemOpen
  \bibfield  {author} {\bibinfo {author} {\bibfnamefont {S.}~\bibnamefont
  {Hayami}}, \bibinfo {author} {\bibfnamefont {S.-Z.}\ \bibnamefont {Lin}},\
  and\ \bibinfo {author} {\bibfnamefont {C.}~\bibnamefont {Batista}},\
  }\bibfield  {title} {\bibinfo {title} {{Bubble and skyrmion crystals in
  frustrated magnets with easy-axis anisotropy}},\ }\href@noop {} {\bibfield
  {journal} {\bibinfo  {journal} {Physical Review B}\ }\textbf {\bibinfo
  {volume} {93}},\ \bibinfo {pages} {184413} (\bibinfo {year}
  {2016})}\BibitemShut {NoStop}%
\bibitem [{\citenamefont {Ozawa}\ \emph {et~al.}(2017)\citenamefont {Ozawa},
  \citenamefont {Hayami},\ and\ \citenamefont {Motome}}]{Ozawa17}%
  \BibitemOpen
  \bibfield  {author} {\bibinfo {author} {\bibfnamefont {R.}~\bibnamefont
  {Ozawa}}, \bibinfo {author} {\bibfnamefont {S.}~\bibnamefont {Hayami}},\ and\
  \bibinfo {author} {\bibfnamefont {Y.}~\bibnamefont {Motome}},\ }\bibfield
  {title} {\bibinfo {title} {{Zero-Field Skyrmions with a High Topological
  Number in Itinerant Magnets}},\ }\href@noop {} {\bibfield  {journal}
  {\bibinfo  {journal} {Physical Review Letters}\ }\textbf {\bibinfo {volume}
  {118}},\ \bibinfo {pages} {147205} (\bibinfo {year} {2017})}\BibitemShut
  {NoStop}%
\bibitem [{\citenamefont {Hayami}\ \emph {et~al.}(2017)\citenamefont {Hayami},
  \citenamefont {Ozawa},\ and\ \citenamefont {Motome}}]{Hayami17}%
  \BibitemOpen
  \bibfield  {author} {\bibinfo {author} {\bibfnamefont {S.}~\bibnamefont
  {Hayami}}, \bibinfo {author} {\bibfnamefont {R.}~\bibnamefont {Ozawa}},\ and\
  \bibinfo {author} {\bibfnamefont {Y.}~\bibnamefont {Motome}},\ }\bibfield
  {title} {\bibinfo {title} {{Effective bilinear-biquadratic model for
  noncoplanar ordering in itinerant magnets}},\ }\href@noop {} {\bibfield
  {journal} {\bibinfo  {journal} {Physical Review B}\ }\textbf {\bibinfo
  {volume} {95}},\ \bibinfo {pages} {224424} (\bibinfo {year}
  {2017})}\BibitemShut {NoStop}%
\bibitem [{\citenamefont {Wang}\ \emph {et~al.}(2020)\citenamefont {Wang},
  \citenamefont {Su}, \citenamefont {Lin},\ and\ \citenamefont
  {Batista}}]{Wang20}%
  \BibitemOpen
  \bibfield  {author} {\bibinfo {author} {\bibfnamefont {Z.}~\bibnamefont
  {Wang}}, \bibinfo {author} {\bibfnamefont {Y.}~\bibnamefont {Su}}, \bibinfo
  {author} {\bibfnamefont {S.-Z.}\ \bibnamefont {Lin}},\ and\ \bibinfo {author}
  {\bibfnamefont {C.}~\bibnamefont {Batista}},\ }\bibfield  {title} {\bibinfo
  {title} {{Skyrmion Crystal from RKKY Interaction Mediated by 2D Electron
  Gas}},\ }\href@noop {} {\bibfield  {journal} {\bibinfo  {journal} {Physical
  Review Letters}\ }\textbf {\bibinfo {volume} {124}},\ \bibinfo {pages}
  {207201} (\bibinfo {year} {2020})}\BibitemShut {NoStop}%
\bibitem [{\citenamefont {Hirschberger}\ \emph {et~al.}(2019)\citenamefont
  {Hirschberger}, \citenamefont {Nakajima}, \citenamefont {Gao}, \citenamefont
  {Peng}, \citenamefont {Kikkawa}, \citenamefont {Kurumaji}, \citenamefont
  {Kriener}, \citenamefont {Yamasaki}, \citenamefont {Sagayama}, \citenamefont
  {Nakao}, \citenamefont {Ohishi}, \citenamefont {Kakurai}, \citenamefont
  {Taguchi}, \citenamefont {Yu}, \citenamefont {Arima},\ and\ \citenamefont
  {Tokura}}]{Hirschberger19}%
  \BibitemOpen
  \bibfield  {author} {\bibinfo {author} {\bibfnamefont {M.}~\bibnamefont
  {Hirschberger}}, \bibinfo {author} {\bibfnamefont {T.}~\bibnamefont
  {Nakajima}}, \bibinfo {author} {\bibfnamefont {S.}~\bibnamefont {Gao}},
  \bibinfo {author} {\bibfnamefont {L.}~\bibnamefont {Peng}}, \bibinfo {author}
  {\bibfnamefont {A.}~\bibnamefont {Kikkawa}}, \bibinfo {author} {\bibfnamefont
  {T.}~\bibnamefont {Kurumaji}}, \bibinfo {author} {\bibfnamefont
  {M.}~\bibnamefont {Kriener}}, \bibinfo {author} {\bibfnamefont
  {Y.}~\bibnamefont {Yamasaki}}, \bibinfo {author} {\bibfnamefont
  {H.}~\bibnamefont {Sagayama}}, \bibinfo {author} {\bibfnamefont
  {H.}~\bibnamefont {Nakao}}, \bibinfo {author} {\bibfnamefont
  {K.}~\bibnamefont {Ohishi}}, \bibinfo {author} {\bibfnamefont
  {K.}~\bibnamefont {Kakurai}}, \bibinfo {author} {\bibfnamefont
  {Y.}~\bibnamefont {Taguchi}}, \bibinfo {author} {\bibfnamefont
  {X.}~\bibnamefont {Yu}}, \bibinfo {author} {\bibfnamefont {T.-h.}\
  \bibnamefont {Arima}},\ and\ \bibinfo {author} {\bibfnamefont
  {Y.}~\bibnamefont {Tokura}},\ }\bibfield  {title} {\bibinfo {title}
  {{Skyrmion phase and competing magnetic orders on a breathing kagom{\'e}
  lattice}},\ }\href@noop {} {\bibfield  {journal} {\bibinfo  {journal} {Nature
  Communications}\ }\textbf {\bibinfo {volume} {10}},\ \bibinfo {pages} {5831}
  (\bibinfo {year} {2019})}\BibitemShut {NoStop}%
\bibitem [{\citenamefont {Khanh}\ \emph {et~al.}(2020)\citenamefont {Khanh},
  \citenamefont {Nakajima}, \citenamefont {Yu}, \citenamefont {Gao},
  \citenamefont {Shibata}, \citenamefont {Hirschberger}, \citenamefont
  {Yamasaki}, \citenamefont {Sagayama}, \citenamefont {Nakao}, \citenamefont
  {Peng}, \citenamefont {Nakajima}, \citenamefont {Takagi}, \citenamefont
  {Arima}, \citenamefont {Tokura},\ and\ \citenamefont {Seki}}]{Khanh20}%
  \BibitemOpen
  \bibfield  {author} {\bibinfo {author} {\bibfnamefont {N.}~\bibnamefont
  {Khanh}}, \bibinfo {author} {\bibfnamefont {T.}~\bibnamefont {Nakajima}},
  \bibinfo {author} {\bibfnamefont {X.}~\bibnamefont {Yu}}, \bibinfo {author}
  {\bibfnamefont {S.}~\bibnamefont {Gao}}, \bibinfo {author} {\bibfnamefont
  {K.}~\bibnamefont {Shibata}}, \bibinfo {author} {\bibfnamefont
  {M.}~\bibnamefont {Hirschberger}}, \bibinfo {author} {\bibfnamefont
  {Y.}~\bibnamefont {Yamasaki}}, \bibinfo {author} {\bibfnamefont
  {H.}~\bibnamefont {Sagayama}}, \bibinfo {author} {\bibfnamefont
  {H.}~\bibnamefont {Nakao}}, \bibinfo {author} {\bibfnamefont
  {L.}~\bibnamefont {Peng}}, \bibinfo {author} {\bibfnamefont {K.}~\bibnamefont
  {Nakajima}}, \bibinfo {author} {\bibfnamefont {R.}~\bibnamefont {Takagi}},
  \bibinfo {author} {\bibfnamefont {T.-h.}\ \bibnamefont {Arima}}, \bibinfo
  {author} {\bibfnamefont {Y.}~\bibnamefont {Tokura}},\ and\ \bibinfo {author}
  {\bibfnamefont {S.}~\bibnamefont {Seki}},\ }\bibfield  {title} {\bibinfo
  {title} {{Nanometric square skyrmion lattice in a centrosymmetric tetragonal
  magnet}},\ }\href@noop {} {\bibfield  {journal} {\bibinfo  {journal} {Nature
  Nanotechnology}\ }\textbf {\bibinfo {volume} {15}},\ \bibinfo {pages}
  {444--449} (\bibinfo {year} {2020})}\BibitemShut {NoStop}%
\bibitem [{\citenamefont {Binz}\ and\ \citenamefont
  {Vishwanath}(2008)}]{Binz08}%
  \BibitemOpen
  \bibfield  {author} {\bibinfo {author} {\bibfnamefont {B.}~\bibnamefont
  {Binz}}\ and\ \bibinfo {author} {\bibfnamefont {A.}~\bibnamefont
  {Vishwanath}},\ }\bibfield  {title} {\bibinfo {title} {{Chirality induced
  anomalous-Hall effect in helical spin crystals}},\ }\href@noop {} {\bibfield
  {journal} {\bibinfo  {journal} {Physica B: Condensed Matter}\ }\textbf
  {\bibinfo {volume} {403}},\ \bibinfo {pages} {1336} (\bibinfo {year}
  {2008})}\BibitemShut {NoStop}%
\bibitem [{\citenamefont {Heinze}\ \emph {et~al.}(2011)\citenamefont {Heinze},
  \citenamefont {von Bergmann}, \citenamefont {Menzel}, \citenamefont {Brede},
  \citenamefont {Kubetzka}, \citenamefont {Wiesendanger}, \citenamefont
  {Bihlmayer},\ and\ \citenamefont {Blügel}}]{Heinze11}%
  \BibitemOpen
  \bibfield  {author} {\bibinfo {author} {\bibfnamefont {S.}~\bibnamefont
  {Heinze}}, \bibinfo {author} {\bibfnamefont {K.}~\bibnamefont {von
  Bergmann}}, \bibinfo {author} {\bibfnamefont {M.}~\bibnamefont {Menzel}},
  \bibinfo {author} {\bibfnamefont {J.}~\bibnamefont {Brede}}, \bibinfo
  {author} {\bibfnamefont {A.}~\bibnamefont {Kubetzka}}, \bibinfo {author}
  {\bibfnamefont {R.}~\bibnamefont {Wiesendanger}}, \bibinfo {author}
  {\bibfnamefont {G.}~\bibnamefont {Bihlmayer}},\ and\ \bibinfo {author}
  {\bibfnamefont {S.}~\bibnamefont {Bl{\"u}gel}},\ }\bibfield  {title} {\bibinfo
  {title} {{Spontaneous atomic-scale magnetic skyrmion lattice in two
  dimensions}},\ }\href@noop {} {\bibfield  {journal} {\bibinfo  {journal}
  {Nature Physics}\ }\textbf {\bibinfo {volume} {7}},\ \bibinfo {pages} {713}
  (\bibinfo {year} {2011})}\BibitemShut {NoStop}%
\bibitem [{\citenamefont {Nagaosa}(2019)}]{Nagaosa19}%
  \BibitemOpen
  \bibfield  {author} {\bibinfo {author} {\bibfnamefont {N.}~\bibnamefont
  {Nagaosa}},\ }\bibfield  {title} {\bibinfo {title} {{Emergent inductor by
  spiral magnets}},\ }\href@noop {} {\bibfield  {journal} {\bibinfo  {journal}
  {Japanese Journal of Applied Physics}\ }\textbf {\bibinfo {volume} {58}},\
  \bibinfo {pages} {120909} (\bibinfo {year} {2019})}\BibitemShut {NoStop}%
\bibitem [{\citenamefont {Sorn}\ \emph {et~al.}(2021)\citenamefont {Sorn},
  \citenamefont {Yang},\ and\ \citenamefont {Paramekanti}}]{Sorn21}%
  \BibitemOpen
  \bibfield  {author} {\bibinfo {author} {\bibfnamefont {S.}~\bibnamefont
  {Sorn}}, \bibinfo {author} {\bibfnamefont {L.}~\bibnamefont {Yang}},\ and\
  \bibinfo {author} {\bibfnamefont {A.}~\bibnamefont {Paramekanti}},\
  }\bibfield  {title} {\bibinfo {title} {{Resonant optical topological Hall
  conductivity from skyrmions}},\ }\href@noop {} {\bibfield  {journal}
  {\bibinfo  {journal} {Physical Review B}\ }\textbf {\bibinfo {volume}
  {104}},\ \bibinfo {pages} {134419} (\bibinfo {year} {2021})}\BibitemShut
  {NoStop}%
\bibitem [{\citenamefont {Kato}\ \emph {et~al.}(2023)\citenamefont {Kato},
  \citenamefont {Okamura}, \citenamefont {Hirschberger}, \citenamefont
  {Tokura},\ and\ \citenamefont {Takahashi}}]{Kato23}%
  \BibitemOpen
  \bibfield  {author} {\bibinfo {author} {\bibfnamefont {Y.}~\bibnamefont
  {Kato}}, \bibinfo {author} {\bibfnamefont {Y.}~\bibnamefont {Okamura}},
  \bibinfo {author} {\bibfnamefont {M.}~\bibnamefont {Hirschberger}}, \bibinfo
  {author} {\bibfnamefont {Y.}~\bibnamefont {Tokura}},\ and\ \bibinfo {author}
  {\bibfnamefont {Y.}~\bibnamefont {Takahashi}},\ }\bibfield  {title} {\bibinfo
  {title} {{Topological magneto-optical effect from skyrmion lattice}},\
  }\href@noop {} {\bibfield  {journal} {\bibinfo  {journal} {Nature
  Communications}\ }\textbf {\bibinfo {volume} {14}},\ \bibinfo {pages} {5416}
  (\bibinfo {year} {2023})}\BibitemShut {NoStop}%
\bibitem [{\citenamefont {Takagi}\ \emph {et~al.}(2022)\citenamefont {Takagi},
  \citenamefont {Matsuyama}, \citenamefont {Ukleev}, \citenamefont {Yu},
  \citenamefont {White}, \citenamefont {Francoual}, \citenamefont {Mardegan},
  \citenamefont {Hayami}, \citenamefont {Saito}, \citenamefont {Kaneko},
  \citenamefont {Ohishi}, \citenamefont {\={O}nuki}, \citenamefont {h.~Arima},
  \citenamefont {Tokura}, \citenamefont {Nakajima},\ and\ \citenamefont
  {Seki}}]{Takagi22}%
  \BibitemOpen
  \bibfield  {author} {\bibinfo {author} {\bibfnamefont {R.}~\bibnamefont
  {Takagi}}, \bibinfo {author} {\bibfnamefont {N.}~\bibnamefont {Matsuyama}},
  \bibinfo {author} {\bibfnamefont {V.}~\bibnamefont {Ukleev}}, \bibinfo
  {author} {\bibfnamefont {L.}~\bibnamefont {Yu}}, \bibinfo {author}
  {\bibfnamefont {J.}~\bibnamefont {White}}, \bibinfo {author} {\bibfnamefont
  {S.}~\bibnamefont {Francoual}}, \bibinfo {author} {\bibfnamefont
  {J.}~\bibnamefont {Mardegan}}, \bibinfo {author} {\bibfnamefont
  {S.}~\bibnamefont {Hayami}}, \bibinfo {author} {\bibfnamefont
  {H.}~\bibnamefont {Saito}}, \bibinfo {author} {\bibfnamefont
  {K.}~\bibnamefont {Kaneko}}, \bibinfo {author} {\bibfnamefont
  {K.}~\bibnamefont {Ohishi}}, \bibinfo {author} {\bibfnamefont
  {Y.}~\bibnamefont {\={O}nuki}}, \bibinfo {author} {\bibfnamefont
  {T.}~\bibnamefont {h.~Arima}}, \bibinfo {author} {\bibfnamefont
  {Y.}~\bibnamefont {Tokura}}, \bibinfo {author} {\bibfnamefont
  {T.}~\bibnamefont {Nakajima}},\ and\ \bibinfo {author} {\bibfnamefont
  {S.}~\bibnamefont {Seki}},\ }\bibfield  {title} {\bibinfo {title} {{Square
  and rhombic lattices of magnetic skyrmions in a centrosymmetric binary
  compound}},\ }\href@noop {} {\bibfield  {journal} {\bibinfo  {journal}
  {Nature Communications}\ }\textbf {\bibinfo {volume} {13}},\ \bibinfo {pages}
  {1472} (\bibinfo {year} {2022})}\BibitemShut {NoStop}%
\bibitem [{\citenamefont {Matsumura}\ \emph {et~al.}(2023)\citenamefont
  {Matsumura}, \citenamefont {Kurauchi}, \citenamefont {Tsukagoshi},
  \citenamefont {Higa}, \citenamefont {Nakao}, \citenamefont {Kakihana},
  \citenamefont {Hedo}, \citenamefont {Nakama},\ and\ \citenamefont
  {\={O}nuki}}]{Matsumura23}%
  \BibitemOpen
  \bibfield  {author} {\bibinfo {author} {\bibfnamefont {T.}~\bibnamefont
  {Matsumura}}, \bibinfo {author} {\bibfnamefont {K.}~\bibnamefont {Kurauchi}},
  \bibinfo {author} {\bibfnamefont {M.}~\bibnamefont {Tsukagoshi}}, \bibinfo
  {author} {\bibfnamefont {N.}~\bibnamefont {Higa}}, \bibinfo {author}
  {\bibfnamefont {H.}~\bibnamefont {Nakao}}, \bibinfo {author} {\bibfnamefont
  {M.}~\bibnamefont {Kakihana}}, \bibinfo {author} {\bibfnamefont
  {M.}~\bibnamefont {Hedo}}, \bibinfo {author} {\bibfnamefont {T.}~\bibnamefont
  {Nakama}},\ and\ \bibinfo {author} {\bibfnamefont {Y.}~\bibnamefont
  {\={O}nuki}},\ }\bibfield  {title} {\bibinfo {title} {{Distorted triangular
  skyrmion lattice in a noncentrosymmetric tetragonal magnet}},\ }\href@noop {}
  {\bibfield  {journal} {\bibinfo  {journal} {arxiv}\ }\textbf {\bibinfo
  {volume} {2306.14767}} (\bibinfo {year} {2023})}\BibitemShut {NoStop}%
\bibitem [{\citenamefont {Zhang}\ \emph {et~al.}(2023)\citenamefont {Zhang},
  \citenamefont {Wang}, \citenamefont {Dahlbom}, \citenamefont {Barros},\ and\
  \citenamefont {Batista}}]{Zhang23}%
  \BibitemOpen
  \bibfield  {author} {\bibinfo {author} {\bibfnamefont {H.}~\bibnamefont
  {Zhang}}, \bibinfo {author} {\bibfnamefont {Z.}~\bibnamefont {Wang}},
  \bibinfo {author} {\bibfnamefont {D.}~\bibnamefont {Dahlbom}}, \bibinfo
  {author} {\bibfnamefont {K.}~\bibnamefont {Barros}},\ and\ \bibinfo {author}
  {\bibfnamefont {C.~D.}\ \bibnamefont {Batista}},\ }\bibfield  {title}
  {\bibinfo {title} {{CP$^2$ skyrmions and skyrmion crystals in realistic
  quantum magnets}},\ }\href@noop {} {\bibfield  {journal} {\bibinfo  {journal}
  {Nature Communications}\ }\textbf {\bibinfo {volume} {14}},\ \bibinfo {pages}
  {3626} (\bibinfo {year} {2023})}\BibitemShut {NoStop}%
\bibitem [{\citenamefont {Hayami}\ and\ \citenamefont
  {Yambe}(2021)}]{Hayami21}%
  \BibitemOpen
  \bibfield  {author} {\bibinfo {author} {\bibfnamefont {S.}~\bibnamefont
  {Hayami}}\ and\ \bibinfo {author} {\bibfnamefont {R.}~\bibnamefont {Yambe}},\
  }\bibfield  {title} {\bibinfo {title} {{Locking of skyrmion cores on a
  centrosymmetric discrete lattice: Onsite versus offsite}},\ }\href@noop {}
  {\bibfield  {journal} {\bibinfo  {journal} {Physical Review Research}\
  }\textbf {\bibinfo {volume} {3}},\ \bibinfo {pages} {043158} (\bibinfo {year}
  {2021})}\BibitemShut {NoStop}%
\bibitem [{\citenamefont {v.~Bergmann}\ \emph {et~al.}(2015)\citenamefont
  {v.~Bergmann}, \citenamefont {Menzel}, \citenamefont {Kubetzka},\ and\
  \citenamefont {Wiesendanger}}]{Bergmann15}%
  \BibitemOpen
  \bibfield  {author} {\bibinfo {author} {\bibfnamefont {K.}~\bibnamefont
  {v.~Bergmann}}, \bibinfo {author} {\bibfnamefont {M.}~\bibnamefont {Menzel}},
  \bibinfo {author} {\bibfnamefont {A.}~\bibnamefont {Kubetzka}},\ and\
  \bibinfo {author} {\bibfnamefont {R.}~\bibnamefont {Wiesendanger}},\
  }\bibfield  {title} {\bibinfo {title} {{Influence of the Local Atom
  Configuration on a Hexagonal Skyrmion Lattice}},\ }\href@noop {} {\bibfield
  {journal} {\bibinfo  {journal} {Nano Letters}\ }\textbf {\bibinfo {volume}
  {15}},\ \bibinfo {pages} {3280--3285} (\bibinfo {year} {2015})}\BibitemShut
  {NoStop}%
\bibitem [{\citenamefont {Wiesendanger}(2016)}]{Wiesendanger16}%
  \BibitemOpen
  \bibfield  {author} {\bibinfo {author} {\bibfnamefont {R.}~\bibnamefont
  {Wiesendanger}},\ }\bibfield  {title} {\bibinfo {title} {{Nanoscale magnetic
  skyrmions in metallic films and multilayers: a new twist for spintronics}},\
  }\href@noop {} {\bibfield  {journal} {\bibinfo  {journal} {Nature Reviews
  Materials}\ }\textbf {\bibinfo {volume} {1}},\ \bibinfo {pages} {16044}
  (\bibinfo {year} {2016})}\BibitemShut {NoStop}%
\bibitem [{\citenamefont {Gutzeit}\ \emph {et~al.}(2023)\citenamefont
  {Gutzeit}, \citenamefont {Drevelow}, \citenamefont {Goerzen}, \citenamefont
  {Haldar},\ and\ \citenamefont {Heinze}}]{Gutzeit23}%
  \BibitemOpen
  \bibfield  {author} {\bibinfo {author} {\bibfnamefont {M.}~\bibnamefont
  {Gutzeit}}, \bibinfo {author} {\bibfnamefont {T.}~\bibnamefont {Drevelow}},
  \bibinfo {author} {\bibfnamefont {M.~A.}\ \bibnamefont {Goerzen}}, \bibinfo
  {author} {\bibfnamefont {S.}~\bibnamefont {Haldar}},\ and\ \bibinfo {author}
  {\bibfnamefont {S.}~\bibnamefont {Heinze}},\ }\bibfield  {title} {\bibinfo
  {title} {{Spontaneous square versus hexagonal nanoscale skyrmion lattices in
  Fe/Ir(111)}},\ }\href@noop {} {\bibfield  {journal} {\bibinfo  {journal}
  {Phys. Rev. B}\ }\textbf {\bibinfo {volume} {108}},\ \bibinfo {pages}
  {L060405} (\bibinfo {year} {2023})}\BibitemShut {NoStop}%
\bibitem [{SI()}]{SI}%
  \BibitemOpen
  \href@noop {} {\bibinfo {title} {{Supplementary Information}}}\BibitemShut
  {NoStop}%
\bibitem [{\citenamefont {Hirschberger}\ \emph {et~al.}(2021)\citenamefont
  {Hirschberger}, \citenamefont {Hayami},\ and\ \citenamefont
  {Tokura}}]{Hirschberger21b}%
  \BibitemOpen
  \bibfield  {author} {\bibinfo {author} {\bibfnamefont {M.}~\bibnamefont
  {Hirschberger}}, \bibinfo {author} {\bibfnamefont {S.}~\bibnamefont
  {Hayami}},\ and\ \bibinfo {author} {\bibfnamefont {Y.}~\bibnamefont
  {Tokura}},\ }\bibfield  {title} {\bibinfo {title} {{Nanometric skyrmion
  lattice from anisotropic exchange interactions in a centrosymmetric host}},\
  }\href@noop {} {\bibfield  {journal} {\bibinfo  {journal} {New Journal of
  Physics}\ }\textbf {\bibinfo {volume} {23}},\ \bibinfo {pages} {023039}
  (\bibinfo {year} {2021})}\BibitemShut {NoStop}%
\bibitem [{\citenamefont {Wilson}\ \emph {et~al.}(1975)\citenamefont {Wilson},
  \citenamefont {Salvo},\ and\ \citenamefont {Mahajan}}]{Wilson1975}%
  \BibitemOpen
  \bibfield  {author} {\bibinfo {author} {\bibfnamefont {J.}~\bibnamefont
  {Wilson}}, \bibinfo {author} {\bibfnamefont {F.~D.}\ \bibnamefont {Salvo}},\
  and\ \bibinfo {author} {\bibfnamefont {S.}~\bibnamefont {Mahajan}},\
  }\bibfield  {title} {\bibinfo {title} {Charge-density waves and superlattices
  in the metallic layered transition metal dichalcogenides},\ }\href@noop {}
  {\bibfield  {journal} {\bibinfo  {journal} {Advances in Physics}\ }\textbf
  {\bibinfo {volume} {24}},\ \bibinfo {pages} {117} (\bibinfo {year}
  {1975})}\BibitemShut {NoStop}%
\bibitem [{\citenamefont {Fleming}\ \emph {et~al.}(1980)\citenamefont
  {Fleming}, \citenamefont {Moncton}, \citenamefont {McWhan},\ and\
  \citenamefont {DiSalvo}}]{Fleming1980}%
  \BibitemOpen
  \bibfield  {author} {\bibinfo {author} {\bibfnamefont {B.~M.}\ \bibnamefont
  {Fleming}}, \bibinfo {author} {\bibfnamefont {D.~E.}\ \bibnamefont
  {Moncton}}, \bibinfo {author} {\bibfnamefont {D.~B.}\ \bibnamefont
  {McWhan}},\ and\ \bibinfo {author} {\bibfnamefont {F.~J.}\ \bibnamefont
  {DiSalvo}},\ }\bibfield  {title} {\bibinfo {title} {{Broken Hexagonal
  Symmetry in the Incommensurate Charge-Density Wave Structure of
  $2H$-TaSe$_2$}},\ }\href@noop {} {\bibfield  {journal} {\bibinfo  {journal}
  {Physical Review Letters}\ }\textbf {\bibinfo {volume} {45}},\ \bibinfo
  {pages} {576} (\bibinfo {year} {1980})}\BibitemShut {NoStop}%
\bibitem [{\citenamefont {Bak}(1982)}]{Bak82}%
  \BibitemOpen
  \bibfield  {author} {\bibinfo {author} {\bibfnamefont {P.}~\bibnamefont
  {Bak}},\ }\bibfield  {title} {\bibinfo {title} {{Commensurate phases,
  incommensurate phases and the devil's staircase}},\ }\href@noop {} {\bibfield
   {journal} {\bibinfo  {journal} {Reports on Progress in Physics}\ }\textbf
  {\bibinfo {volume} {45}},\ \bibinfo {pages} {587} (\bibinfo {year}
  {1982})}\BibitemShut {NoStop}%
\bibitem [{\citenamefont {Hase}\ \emph {et~al.}(1993)\citenamefont {Hase},
  \citenamefont {Terasaki},\ and\ \citenamefont {Uchinokura}}]{Hase93}%
  \BibitemOpen
  \bibfield  {author} {\bibinfo {author} {\bibfnamefont {M.}~\bibnamefont
  {Hase}}, \bibinfo {author} {\bibfnamefont {I.}~\bibnamefont {Terasaki}},\
  and\ \bibinfo {author} {\bibfnamefont {K.}~\bibnamefont {Uchinokura}},\
  }\bibfield  {title} {\bibinfo {title} {{Observation of the spin-Peierls
  transition in linear Cu$^{+}$ (spin-$1/2$) chains in an inorganic compound
  CuGeO$_3$}},\ }\href@noop {} {\bibfield  {journal} {\bibinfo  {journal}
  {Physical Review Letters}\ }\textbf {\bibinfo {volume} {70}},\ \bibinfo
  {pages} {3651} (\bibinfo {year} {1993})}\BibitemShut {NoStop}%
\bibitem [{\citenamefont {Kiryukhin}\ \emph {et~al.}(1995)\citenamefont
  {Kiryukhin}, \citenamefont {Keimer},\ and\ \citenamefont
  {Moncton}}]{Kiryukhin95a}%
  \BibitemOpen
  \bibfield  {author} {\bibinfo {author} {\bibfnamefont {V.}~\bibnamefont
  {Kiryukhin}}, \bibinfo {author} {\bibfnamefont {B.}~\bibnamefont {Keimer}},\
  and\ \bibinfo {author} {\bibfnamefont {D.~E.}\ \bibnamefont {Moncton}},\
  }\bibfield  {title} {\bibinfo {title} {{Direct Observation of a Magnetic
  Field Induced Commensurate-Incommensurate Transition in a Spin-Peierls
  System}},\ }\href@noop {} {\bibfield  {journal} {\bibinfo  {journal}
  {Physical Review Letters}\ }\textbf {\bibinfo {volume} {74}},\ \bibinfo
  {pages} {1669} (\bibinfo {year} {1995})}\BibitemShut {NoStop}%
\bibitem [{\citenamefont {Kiryukhin}\ and\ \citenamefont
  {Keimer}(1995)}]{Kiryukhin95b}%
  \BibitemOpen
  \bibfield  {author} {\bibinfo {author} {\bibfnamefont {V.}~\bibnamefont
  {Kiryukhin}}\ and\ \bibinfo {author} {\bibfnamefont {B.}~\bibnamefont
  {Keimer}},\ }\bibfield  {title} {\bibinfo {title} {{Incommensurate lattice
  modulation in the spin-Peierls system CuGeO$_3$}},\ }\href@noop {} {\bibfield
   {journal} {\bibinfo  {journal} {Physical Review B}\ }\textbf {\bibinfo
  {volume} {52}},\ \bibinfo {pages} {R704(R)} (\bibinfo {year}
  {1995})}\BibitemShut {NoStop}%
\bibitem [{\citenamefont {Kiryukhin}\ \emph {et~al.}(1996)\citenamefont
  {Kiryukhin}, \citenamefont {Keimer}, \citenamefont {Hill},\ and\
  \citenamefont {Vigliante}}]{Kiryukhin96}%
  \BibitemOpen
  \bibfield  {author} {\bibinfo {author} {\bibfnamefont {V.}~\bibnamefont
  {Kiryukhin}}, \bibinfo {author} {\bibfnamefont {B.}~\bibnamefont {Keimer}},
  \bibinfo {author} {\bibfnamefont {J.}~\bibnamefont {Hill}},\ and\ \bibinfo
  {author} {\bibfnamefont {A.}~\bibnamefont {Vigliante}},\ }\bibfield  {title}
  {\bibinfo {title} {{Soliton Lattice in Pure and Diluted CuGeO$_3$}},\
  }\href@noop {} {\bibfield  {journal} {\bibinfo  {journal} {Physical Review
  Letters}\ }\textbf {\bibinfo {volume} {76}},\ \bibinfo {pages} {4608}
  (\bibinfo {year} {1996})}\BibitemShut {NoStop}%
\bibitem [{\citenamefont {Leonov}\ and\ \citenamefont
  {Mostovoy}(2015)}]{Leonov15}%
  \BibitemOpen
  \bibfield  {author} {\bibinfo {author} {\bibfnamefont {A.~O.}\ \bibnamefont
  {Leonov}}\ and\ \bibinfo {author} {\bibfnamefont {M.}~\bibnamefont
  {Mostovoy}},\ }\bibfield  {title} {\bibinfo {title} {{Multiply periodic
  states and isolated skyrmions in an anisotropic frustrated magnet}},\
  }\href@noop {} {\bibfield  {journal} {\bibinfo  {journal} {Nature
  Communications}\ }\textbf {\bibinfo {volume} {6}},\ \bibinfo {pages} {8275}
  (\bibinfo {year} {2015})}\BibitemShut {NoStop}%
\bibitem [{\citenamefont {Koehler}(1972)}]{Koehler72}%
  \BibitemOpen
  \bibfield  {author} {\bibinfo {author} {\bibfnamefont {W.}~\bibnamefont
  {Koehler}},\ }\href@noop {} {\emph {\bibinfo {title} {{Magnetic Structures of
  Rare Earth Metals and Alloys}}}},\ edited by\ \bibinfo {editor}
  {\bibfnamefont {R.}~\bibnamefont {Elliott}}\ (\bibinfo  {publisher} {Springer
  New York},\ \bibinfo {year} {1972})\BibitemShut {NoStop}%
\bibitem [{\citenamefont {Jensen}\ and\ \citenamefont
  {Mackintosh}(1991)}]{Jensen91}%
  \BibitemOpen
  \bibfield  {author} {\bibinfo {author} {\bibfnamefont {J.}~\bibnamefont
  {Jensen}}\ and\ \bibinfo {author} {\bibfnamefont {A.~R.}\ \bibnamefont
  {Mackintosh}},\ }\href@noop {} {\emph {\bibinfo {title} {{Rare Earth
  Magnetism: Structures and Excitations}}}}\ (\bibinfo  {publisher} {Clarendon
  Press},\ \bibinfo {year} {1991})\BibitemShut {NoStop}%
\bibitem [{\citenamefont {Venturini}\ \emph {et~al.}(1996)\citenamefont
  {Venturini}, \citenamefont {Fruchart},\ and\ \citenamefont
  {Malaman}}]{Venturini96}%
  \BibitemOpen
  \bibfield  {author} {\bibinfo {author} {\bibfnamefont {G.}~\bibnamefont
  {Venturini}}, \bibinfo {author} {\bibfnamefont {D.}~\bibnamefont
  {Fruchart}},\ and\ \bibinfo {author} {\bibfnamefont {B.}~\bibnamefont
  {Malaman}},\ }\bibfield  {title} {\bibinfo {title} {{Incommensurate magnetic
  structures of $R$Mn$_6$Sn$_6$ ($R$ = Sc, Y, Lu) compounds from neutron
  diffraction study}},\ }\href@noop {} {\bibfield  {journal} {\bibinfo
  {journal} {Journal of Alloys and Compounds}\ }\textbf {\bibinfo {volume}
  {236}},\ \bibinfo {pages} {102} (\bibinfo {year} {1996})}\BibitemShut
  {NoStop}%
\bibitem [{\citenamefont {Gardner}\ \emph {et~al.}(2010)\citenamefont
  {Gardner}, \citenamefont {Gingras},\ and\ \citenamefont
  {Greedan}}]{Gardner10}%
  \BibitemOpen
  \bibfield  {author} {\bibinfo {author} {\bibfnamefont {J.}~\bibnamefont
  {Gardner}}, \bibinfo {author} {\bibfnamefont {M.}~\bibnamefont {Gingras}},\
  and\ \bibinfo {author} {\bibfnamefont {J.}~\bibnamefont {Greedan}},\
  }\bibfield  {title} {\bibinfo {title} {{Magnetic pyrochlore oxides}},\
  }\href@noop {} {\bibfield  {journal} {\bibinfo  {journal} {Reviews of Modern
  Physics}\ }\textbf {\bibinfo {volume} {82}},\ \bibinfo {pages} {53} (\bibinfo
  {year} {2010})}\BibitemShut {NoStop}%
\bibitem [{\citenamefont {Takagi}\ \emph {et~al.}(2023)\citenamefont {Takagi},
  \citenamefont {Takagi}, \citenamefont {Minami}, \citenamefont {Nomoto},
  \citenamefont {Ohishi}, \citenamefont {Suzuki}, \citenamefont {Yanagi},
  \citenamefont {Hirayama}, \citenamefont {Khanh}, \citenamefont {Karube},
  \citenamefont {Saito}, \citenamefont {Hashizume}, \citenamefont {Kiyanagi},
  \citenamefont {Tokura}, \citenamefont {Arita}, \citenamefont {Nakajima},\
  and\ \citenamefont {Seki}}]{Takagi23}%
  \BibitemOpen
  \bibfield  {author} {\bibinfo {author} {\bibfnamefont {H.}~\bibnamefont
  {Takagi}}, \bibinfo {author} {\bibfnamefont {R.}~\bibnamefont {Takagi}},
  \bibinfo {author} {\bibfnamefont {S.}~\bibnamefont {Minami}}, \bibinfo
  {author} {\bibfnamefont {T.}~\bibnamefont {Nomoto}}, \bibinfo {author}
  {\bibfnamefont {K.}~\bibnamefont {Ohishi}}, \bibinfo {author} {\bibfnamefont
  {M.-T.}\ \bibnamefont {Suzuki}}, \bibinfo {author} {\bibfnamefont
  {Y.}~\bibnamefont {Yanagi}}, \bibinfo {author} {\bibfnamefont
  {M.}~\bibnamefont {Hirayama}}, \bibinfo {author} {\bibfnamefont {N.~D.}\
  \bibnamefont {Khanh}}, \bibinfo {author} {\bibfnamefont {K.}~\bibnamefont
  {Karube}}, \bibinfo {author} {\bibfnamefont {H.}~\bibnamefont {Saito}},
  \bibinfo {author} {\bibfnamefont {D.}~\bibnamefont {Hashizume}}, \bibinfo
  {author} {\bibfnamefont {R.}~\bibnamefont {Kiyanagi}}, \bibinfo {author}
  {\bibfnamefont {Y.}~\bibnamefont {Tokura}}, \bibinfo {author} {\bibfnamefont
  {R.}~\bibnamefont {Arita}}, \bibinfo {author} {\bibfnamefont
  {T.}~\bibnamefont {Nakajima}},\ and\ \bibinfo {author} {\bibfnamefont
  {S.}~\bibnamefont {Seki}},\ }\bibfield  {title} {\bibinfo {title}
  {{Spontaneous topological Hall effect induced by non-coplanar
  antiferromagnetic order in intercalated van der Waals materials}},\
  }\href@noop {} {\bibfield  {journal} {\bibinfo  {journal} {Nature Physics}\
  }\textbf {\bibinfo {volume} {19}},\ \bibinfo {pages} {961--968} (\bibinfo
  {year} {2023})}\BibitemShut {NoStop}%
\bibitem [{\citenamefont {Bak}\ and\ \citenamefont {Jensen}(1980)}]{Bak80}%
  \BibitemOpen
  \bibfield  {author} {\bibinfo {author} {\bibfnamefont {P.}~\bibnamefont
  {Bak}}\ and\ \bibinfo {author} {\bibfnamefont {M.}~\bibnamefont {Jensen}},\
  }\bibfield  {title} {\bibinfo {title} {{Theory of helical magnetic structures
  and phase transitions in MnSi and FeGe}},\ }\href@noop {} {\bibfield
  {journal} {\bibinfo  {journal} {Journal of Physics C: Solid State Physics}\
  }\textbf {\bibinfo {volume} {13}},\ \bibinfo {pages} {L881} (\bibinfo {year}
  {1980})}\BibitemShut {NoStop}%
\bibitem [{\citenamefont {Butykai}\ \emph {et~al.}(2022)\citenamefont
  {Butykai}, \citenamefont {Geirhos}, \citenamefont {Szaller}, \citenamefont
  {Kiss}, \citenamefont {Balogh}, \citenamefont {Azhar}, \citenamefont {Garst},
  \citenamefont {DeBeer-Schmitt}, \citenamefont {Waki}, \citenamefont {Tabata},
  \citenamefont {Nakamura}, \citenamefont {K{\'e}zsm{\'a}rki},\ and\
  \citenamefont {Bord{\'a}cs}}]{Butykai22}%
  \BibitemOpen
  \bibfield  {author} {\bibinfo {author} {\bibfnamefont {{\'A}.}~\bibnamefont
  {Butykai}}, \bibinfo {author} {\bibfnamefont {K.}~\bibnamefont {Geirhos}},
  \bibinfo {author} {\bibfnamefont {D.}~\bibnamefont {Szaller}}, \bibinfo
  {author} {\bibfnamefont {L.~F.}\ \bibnamefont {Kiss}}, \bibinfo {author}
  {\bibfnamefont {L.}~\bibnamefont {Balogh}}, \bibinfo {author} {\bibfnamefont
  {M.}~\bibnamefont {Azhar}}, \bibinfo {author} {\bibfnamefont
  {M.}~\bibnamefont {Garst}}, \bibinfo {author} {\bibfnamefont
  {L.}~\bibnamefont {DeBeer-Schmitt}}, \bibinfo {author} {\bibfnamefont
  {T.}~\bibnamefont {Waki}}, \bibinfo {author} {\bibfnamefont {Y.}~\bibnamefont
  {Tabata}}, \bibinfo {author} {\bibfnamefont {H.}~\bibnamefont {Nakamura}},
  \bibinfo {author} {\bibfnamefont {I.}~\bibnamefont {K{\'e}zsm{\'a}rki}},\
  and\ \bibinfo {author} {\bibfnamefont {S.}~\bibnamefont {Bord{\'a}cs}},\
  }\bibfield  {title} {\bibinfo {title} {{Squeezing the periodicity of
  N{\'e}el-type magnetic modulations by enhanced Dzyaloshinskii-Moriya
  interaction of $4d$ electrons}},\ }\href@noop {} {\bibfield  {journal}
  {\bibinfo  {journal} {npj Quantum Materials}\ }\textbf {\bibinfo {volume}
  {7}},\ \bibinfo {pages} {26} (\bibinfo {year} {2022})}\BibitemShut {NoStop}%
\bibitem [{\citenamefont {Snyder}(1989)}]{Snyder89}%
  \BibitemOpen
  \bibfield  {author} {\bibinfo {author} {\bibfnamefont {J.~P.}\ \bibnamefont
  {Snyder}},\ }\bibfield  {title} {\bibinfo {title} {{An Album of Map
  Projections}},\ }\href@noop {} {\bibfield  {journal} {\bibinfo  {journal}
  {US. Geological Survey}\ }\textbf {\bibinfo {volume} {1453}},\ \bibinfo
  {pages} {234} (\bibinfo {year} {1989})}\BibitemShut {NoStop}%
\bibitem [{\citenamefont {Izumi}\ and\ \citenamefont {Momma}(2007)}]{Izumi07}%
  \BibitemOpen
  \bibfield  {author} {\bibinfo {author} {\bibfnamefont {F.}~\bibnamefont
  {Izumi}}\ and\ \bibinfo {author} {\bibfnamefont {K.}~\bibnamefont {Momma}},\
  }\bibfield  {title} {\bibinfo {title} {{Three-Dimensional Visualization in
  Powder Diffraction}},\ }\href@noop {} {\bibfield  {journal} {\bibinfo
  {journal} {Solid State Phenomena}\ }\textbf {\bibinfo {volume} {130}},\
  \bibinfo {pages} {15} (\bibinfo {year} {2007})}\BibitemShut {NoStop}%
\bibitem [{\citenamefont {N{\'a}fr{\'a}di}\ \emph {et~al.}(2008)\citenamefont
  {N{\'a}fr{\'a}di}, \citenamefont {Ga{\'a}l}, \citenamefont {Sienkiewicz},
  \citenamefont {Feh{\'e}r},\ and\ \citenamefont {Forr{\'o}}}]{Nafradi08}%
  \BibitemOpen
  \bibfield  {author} {\bibinfo {author} {\bibfnamefont {B.}~\bibnamefont
  {N{\'a}fr{\'a}di}}, \bibinfo {author} {\bibfnamefont {R.}~\bibnamefont
  {Ga{\'a}l}}, \bibinfo {author} {\bibfnamefont {A.}~\bibnamefont
  {Sienkiewicz}}, \bibinfo {author} {\bibfnamefont {T.}~\bibnamefont
  {Feh{\'e}r}},\ and\ \bibinfo {author} {\bibfnamefont {L.}~\bibnamefont
  {Forr{\'o}}},\ }\bibfield  {title} {\bibinfo {title} {{Continuous-wave
  far-infrared ESR spectrometer for high-pressure measurements}},\ }\href@noop
  {} {\bibfield  {journal} {\bibinfo  {journal} {Journal of Magnetic
  Resonance}\ }\textbf {\bibinfo {volume} {195}},\ \bibinfo {pages} {206}
  (\bibinfo {year} {2008})}\BibitemShut {NoStop}%
\bibitem [{\citenamefont {Ehlers}\ \emph {et~al.}(2017)\citenamefont {Ehlers},
  \citenamefont {Stasinopoulos}, \citenamefont {K{\'e}zsm{\'a}rki},
  \citenamefont {Feh{\'e}r}, \citenamefont {Tsurkan}, \citenamefont {Krug~von
  Nidda}, \citenamefont {Grundler},\ and\ \citenamefont {Loidl}}]{Ehlers17}%
  \BibitemOpen
  \bibfield  {author} {\bibinfo {author} {\bibfnamefont {D.}~\bibnamefont
  {Ehlers}}, \bibinfo {author} {\bibfnamefont {I.}~\bibnamefont
  {Stasinopoulos}}, \bibinfo {author} {\bibfnamefont {I.}~\bibnamefont
  {K{\'e}zsm{\'a}rki}}, \bibinfo {author} {\bibfnamefont {T.}~\bibnamefont
  {Feh{\'e}r}}, \bibinfo {author} {\bibfnamefont {V.}~\bibnamefont {Tsurkan}},
  \bibinfo {author} {\bibfnamefont {H.-A.}\ \bibnamefont {Krug~von Nidda}},
  \bibinfo {author} {\bibfnamefont {D.}~\bibnamefont {Grundler}},\ and\
  \bibinfo {author} {\bibfnamefont {A.}~\bibnamefont {Loidl}},\ }\bibfield
  {title} {\bibinfo {title} {{Exchange anisotropy in the skyrmion host
  GaV$_4$S$_8$}},\ }\href@noop {} {\bibfield  {journal} {\bibinfo  {journal}
  {Journal of Physics: Condensed Matter}\ }\textbf {\bibinfo {volume} {29}},\
  \bibinfo {pages} {065803} (\bibinfo {year} {2017})}\BibitemShut {NoStop}%
\bibitem [{\citenamefont {White}(2007)}]{White07}%
  \BibitemOpen
  \bibfield  {author} {\bibinfo {author} {\bibfnamefont {R.}~\bibnamefont
  {White}},\ }\href@noop {} {\emph {\bibinfo {title} {{Quantum Theory of
  Magnetism}}}}\ (\bibinfo  {publisher} {Springer Berlin, Heidelberg},\
  \bibinfo {year} {2007})\BibitemShut {NoStop}%
\bibitem [{\citenamefont {Gao}\ \emph {et~al.}(2019)\citenamefont {Gao},
  \citenamefont {Hirschberger}, \citenamefont {Zaharko}, \citenamefont
  {Nakajima}, \citenamefont {Kurumaji}, \citenamefont {Kikkawa}, \citenamefont
  {Shiogai}, \citenamefont {Tsukazaki}, \citenamefont {Kimura}, \citenamefont
  {Awaji}, \citenamefont {Taguchi}, \citenamefont {Arima},\ and\ \citenamefont
  {Tokura}}]{Gao19}%
  \BibitemOpen
  \bibfield  {author} {\bibinfo {author} {\bibfnamefont {S.}~\bibnamefont
  {Gao}}, \bibinfo {author} {\bibfnamefont {M.}~\bibnamefont {Hirschberger}},
  \bibinfo {author} {\bibfnamefont {O.}~\bibnamefont {Zaharko}}, \bibinfo
  {author} {\bibfnamefont {T.}~\bibnamefont {Nakajima}}, \bibinfo {author}
  {\bibfnamefont {T.}~\bibnamefont {Kurumaji}}, \bibinfo {author}
  {\bibfnamefont {A.}~\bibnamefont {Kikkawa}}, \bibinfo {author} {\bibfnamefont
  {J.}~\bibnamefont {Shiogai}}, \bibinfo {author} {\bibfnamefont
  {A.}~\bibnamefont {Tsukazaki}}, \bibinfo {author} {\bibfnamefont
  {S.}~\bibnamefont {Kimura}}, \bibinfo {author} {\bibfnamefont
  {S.}~\bibnamefont {Awaji}}, \bibinfo {author} {\bibfnamefont
  {Y.}~\bibnamefont {Taguchi}}, \bibinfo {author} {\bibfnamefont {T.-h.}\
  \bibnamefont {Arima}},\ and\ \bibinfo {author} {\bibfnamefont
  {Y.}~\bibnamefont {Tokura}},\ }\bibfield  {title} {\bibinfo {title}
  {{Ordering phenomena of spin trimers accompanied by a large geometrical Hall
  effect}},\ }\href@noop {} {\bibfield  {journal} {\bibinfo  {journal}
  {Physical Review B}\ }\textbf {\bibinfo {volume} {100}},\ \bibinfo {pages}
  {241115} (\bibinfo {year} {2019})}\BibitemShut {NoStop}%
\bibitem [{\citenamefont {Nakamura}\ \emph {et~al.}(2023)\citenamefont
  {Nakamura}, \citenamefont {Kabeya}, \citenamefont {Kobayashi}, \citenamefont
  {Araki}, \citenamefont {Katoh},\ and\ \citenamefont {Ochiai}}]{Nakamura23}%
  \BibitemOpen
  \bibfield  {author} {\bibinfo {author} {\bibfnamefont {S.}~\bibnamefont
  {Nakamura}}, \bibinfo {author} {\bibfnamefont {N.}~\bibnamefont {Kabeya}},
  \bibinfo {author} {\bibfnamefont {M.}~\bibnamefont {Kobayashi}}, \bibinfo
  {author} {\bibfnamefont {K.}~\bibnamefont {Araki}}, \bibinfo {author}
  {\bibfnamefont {K.}~\bibnamefont {Katoh}},\ and\ \bibinfo {author}
  {\bibfnamefont {A.}~\bibnamefont {Ochiai}},\ }\bibfield  {title} {\bibinfo
  {title} {{Magnetic phases of the frustrated ferromagnetic spin-trimer system
  ${\mathrm{Gd}}_{3}{\mathrm{Ru}}_{4}{\mathrm{Al}}_{12}$ with a distorted
  kagome lattice structure}},\ }\href@noop {} {\bibfield  {journal} {\bibinfo
  {journal} {Phys. Rev. B}\ }\textbf {\bibinfo {volume} {107}},\ \bibinfo
  {pages} {014422} (\bibinfo {year} {2023})}\BibitemShut {NoStop}%
\end{thebibliography}
\end{document}